\documentclass[aps,prd,onecolumn,groupedaddress,showpacs,nofootinbib,amssymb]{revtex4-2}
\usepackage[dvips]{graphicx}
\usepackage{amssymb}
\usepackage{amsmath}
\usepackage{graphicx,color}
\usepackage{amsfonts}
\usepackage{bm}
\usepackage{cancel}
\usepackage{comment}
\usepackage{hyperref}
\usepackage{ulem}

\newcommand{\e}{\mathrm{e}}

\allowdisplaybreaks[4]

\begin{document}

\tolerance=5000
\title{Time dependent black holes and gravitational wave in Einstein--Gauss-Bonnet theory with two scalar fields}
\author{G.~G.~L.~Nashed}\email{nashed@bue.edu.eg}
\affiliation{ Center for Theoretical Physics, British University of Egypt,
Sherouk City 11837, Egypt\\Center for Space Research, North-West University, Potchefstroom 2520, South Africa}

\begin{abstract}
This paper explores time-varying black holes within the framework of the Einstein-Gauss-Bonnet theory with two scalar fields, examining the propagation of gravitational waves (GW).  In reconstructed models,  ghosts frequently emerge but can be eliminated by applying certain constraints.  We investigate the behavior of high-frequency gravitational waves by examining the effects of varying Gauss-Bonnet coupling during their propagation. The speed of transmission is altered by the coupling in the creation of black holes.
The speed of gravitational waves varies as they enter a black hole compared to when they exit.
\end{abstract}

\maketitle

\section{Introduction}\label{SecI}

{ Black holes reside within the rapidly growing universe.  Specifically, in our growing universe, there are no asymptotically flat black holes. Therefore, studying the physics and thermodynamics of black holes within a growing cosmological environment is essential.  Thus, in recent years, there has been much research on black holes that are asymptotic to the expanding universe, referred to as ``Cosmological Black Holes''.  These black holes raise various questions as: What effects does the cosmic expansion have on the local physics of black holes in the entire cosmic epochs?  What impact does the universe's content have on the black hole? What changes should be made to the physics of black holes in light of the expanding universe? In an expanding universe, how should the definitions of black hole horizon, singularities, mass, and thermodynamics be altered? McVittie's solution is one of the earlier studies that explains black holes within the Friedmann-Robertson-Walker (FRW) universe \cite{McVittie:1933zz}. Subsequently, solutions such as those developed by Einstein and Strauss \cite{Einstein:1945id}, Lemaitre-Tolman-Bondi (LTB) \cite{Tolman:1934za,Bondi:1947fta,Lemaitre:1933gd}, and Vaidya \cite{Vaidya:1977zza} were introduced.  In this study, the significant aspect is the reconsideration of boundaries using indigenous ideas instead of asymptotically flat circumstances, a concept proposed as trapping horizon by Hayward \cite{Hayward:1993wb} and as dynamical horizon \cite{Ashtekar:2002ag}. In addition to the LTB metric's dynamic nature, the FLRW metric can be represented as a background and is a specific instance of the LTB metric. By utilizing the characteristics of the LTB metric, it is possible to create a cosmological black hole \citep{Firouzjaee:2010whc}, with both its singularity and horizon emerging as it collapses \citep{Firouzjaee:2011dn}. References \cite{Vanzo:2011wq, Faraoni:2015ula} provide useful overviews of different types of horizons such as event, Killing, apparent, trapping, isolated, and dynamical horizons. Within modified theories like $f(R)$, $f(G)$, etc. it is difficult to derive time dependent solution \cite{Nashed:2020mnp,Nojiri:2023qgd}. However, for theories coupled with scalar fields it can be applicable to derive time-dependent solution \cite{Nojiri:2020blr}.}

Einstein-Gauss-Bonnet (EGB) theories are significant because they emerge as ultraviolet corrections to the minimally coupled scalar field theory \cite{Codello:2015mba}. The EGB theory is a promising candidate for explaining the inflationary epoch \cite{Nojiri:2010wj, Nojiri:2017ncd, Hwang:2005hb,ElHanafy:2014efn, Nojiri:2006je, Nojiri:2005vv, Satoh:2007gn, Yi:2018gse, Guo:2009uk, Jiang:2013gza, Kanti:2015pda, vandeBruck:2017voa, Kanti:1998jd, Pozdeeva:2020apf, Pozdeeva:2021iwc, Chervon:2019sey, Koh:2014bka,Nashed:2015pga, Bayarsaikhan:2020jww,DeLaurentis:2015fea, Nozari:2017rta, Odintsov:2018zhw, Kawai:1998ab, Yi:2018dhl, vandeBruck:2016xvt, Kleihaus:2019rbg, Bakopoulos:2019tvc, Maeda:2011zn, Bakopoulos:2020dfg, Ai:2020peo, Odintsov:2020xji, Oikonomou:2020sij, Venikoudis:2021irr,Odintsov:2020zkl, Odintsov:2020mkz, Kong:2021qiu, Easther:1996yd, Nashed:2018qag,Antoniadis:1993jc, Antoniadis:1990uu, Rashidi:2020wwg, Odintsov:2023aaw,  Kanti:1995vq, Kanti:1997br, Easson:2020mpq,Nashed:2022zyi, Odintsov:2023lbb,Nashed:2018piz,Oikonomou:2022xoq,Odintsov:2023weg, Nojiri:2023jtf,Nashed:2001im,TerenteDiaz:2023kgc,Kawai:2023nqs,Kawai:2021edk,Kawai:2017kqt,Choudhury:2023kam}. One of its key features is its ability to generate a desirable blue-tilted tensor spectral index. This property is crucial because it enables the stochastic gravitational waves produced during an EGB  inflationary era to be potentially detected in future gravitational wave experiments. Recent research \cite{Nojiri:2020blr} demonstrated that an arbitrary spherically symmetric spacetime can be formulated by coupling general relativity  with two scalar fields, even when the spacetime itself is dynamic. Research has shown that ghost degrees of freedom can emerge in certain models. In classical theory, these ghosts have kinetic energy that is unbounded from below, and in quantum theory, they produce negative norm states, a phenomenon also observed in quantum chromodynamics \cite{Kugo:1979gm}. This is significant because it conflicts with the Copenhagen interpretation of quantum theory. The presence of ghost degrees of freedom poses a challenge to physical reality. Recently, an extension of the formulation in Ref. \cite{Nojiri:2020blr} has led to proposed models describing static wormholes \cite{Nojiri:2023dvf} and gravastars \cite{Nojiri:2023zlp}. Furthermore, these models have been investigated for the propagation of gravitational waves in spherically symmetric spacetimes \cite{Nojiri:2023mbo}. Gravastars offer an alternative to conventional black holes and could potentially form through Bose-Einstein condensation \cite{Mazur:2004fk}. In previous works \cite{Nojiri:2020blr,Nojiri:2023dvf}, constraints were imposed on the scalar fields to prevent the emergence of ghost degrees of freedom. The constraints used resembled those employed in mimetic gravity \cite{Chamseddine:2013kea}. Consequently, the scalar fields lose their dynamical behavior, effectively prohibiting fluctuations and preventing the propagation of these fields. In mimetic gravity, the constraints employed mirror those used, causing the scalar fields to lose their dynamic behavior. This effectively prevents fluctuations and hinders the propagation of these fields. In the context of the original mimetic gravity approach, this scenario closely resembles a situation where effective dark matter emerges. Unlike standard dust, this effective dark matter exhibits vanishing pressure, yet it remains non-dynamical and does not undergo gravitational collapse.

In this study, we examine an extension of the model presented in \cite{Nojiri:2020blr}, which involves coupling the Gauss-Bonnet invariant with two scalar fields because of the substantial phenomenological implications of EGB.
Furthermore, the Gauss-Bonnet term emerges as $\alpha$ correction from string theory to Einstein's gravity, with the parameter $1/\alpha$ representing the string tension.  We designate this model, which encompasses both scalar fields and the Gauss-Bonnet invariant, as EGB gravity. Moreover,  we show that within the present study, it is impossible to construct a model in which the propagation speed of gravitational waves coincides with that of light in dynamical spherically symmetric spacetimes.

The arrangement of this study is as follows: In Section~\ref{SecII}, we present in a brief scenario the EGB theory coupled with two scalar fields.
In Section~\ref{sec2}, we apply the field of EGB coupled with two scalar fields to a time-dependent spherically symmetric spacetime and obtain a system of differential equations. This system is in a closed form, which enabled us to derive explicitly the forms of the coefficients of the scalar field besides the potential. In Section \ref{sec3} we present three time-dependent black holes and derive their relevant coefficient of scalar field as well as the potential.  In Section \ref{sec4} we discuss the gravitational wave by presenting the perturbed form of Einstein-Gauss-Bonnet theory coupled with two scalar fields. In the final section, we investigate the results contained in this study.

\section{Ghost-free }\label{SecII}

  According to the study presented in~\cite{Nojiri:2020blr} it is shown that any spherically symmetric spacetime, whether dynamic or static, can be constructed utilizing Einstein's theory in conjunction with having two scalar fields. The theory \cite{Nojiri:2020blr}, however, involves ghost\footnote{{ In classical terms, the ghost mode has negative kinetic energy which, in quantum theory, results in the formation of negative norm states. Therefore, if a ghost mode is present, it shows that the model lacks physical consistency.}}. The present work develops solutions having no ghosts by presenting limits via the Lagrange multiplier, analogous to the limits found in mimetic theory ~\cite{Chamseddine:2013kea}. Moreover, to account for variations in the propagation speed of gravitational waves, We present the connection between two scalar fields and the Gauss-Bonnet term, following the approach outlined in the references \cite{Nojiri:2005vv,Nojiri:2006je}.

We start with EGB gravity coupled to two scalar fields, denoted as $\varphi$ and $\psi$. The action for this system is given by:
\begin{align}
\label{I88}
S_{\varphi\psi} = \int d^4 x \sqrt{-g} & \left\{ \frac{R}{2\kappa^2}
 - \frac{1}{2} \beta(\varphi,\psi) \partial_\mu \varphi \partial^\mu \varphi
 - \sigma(\varphi,\psi) \partial_\mu \varphi \partial^\mu \psi- \frac{1}{2} \alpha (\varphi,\psi) \partial_\mu \psi \partial^\mu \psi
 - \Phi (\varphi,\psi) - \zeta(\varphi, \psi) \mathcal{G} + \mathcal{L}_\mathrm{matter} \right\}\, .
\end{align}
Here, $\Phi(\varphi,\psi)$  represents the potential which is a function of  $\varphi$ and $\psi$, and $\zeta(\varphi, \psi)$ is coupling function of the Gauss-Bonnet which is a function of $\varphi$ and $\psi$. Here  $\mathcal{G}$ represents Gauss-Bonnet invariant term given by
\begin{align}
\label{eq:GB}
\mathcal{G} = R^2-4R_{\alpha \beta}R^{\alpha \beta}+R_{\alpha \beta \rho \sigma}R^{\alpha \beta \rho \sigma}\, ,
\end{align}
that is a topological density.  In four dimensions Gauss-Bonnet is a total derivative that yields the Euler number.

Varying the action of Eq.~(\ref{I88}) regarding
metric $g_{\mu\nu}$, gives
\begin{align}
\label{gb4bD44}
0= &\, \frac{1}{2\kappa^2}\left(- R_{\mu\nu} + \frac{1}{2} g_{\mu\nu} R\right)  + \frac{1}{2} g_{\mu\nu} \left\{
 - \frac{1}{2} \beta(\varphi,\psi) \partial_\rho \varphi \partial^\rho \varphi
 - \sigma(\varphi,\psi) \partial_\rho \varphi \partial^\rho \psi
 - \frac{1}{2} \alpha (\varphi,\psi) \partial_\rho \psi \partial^\rho \psi - \Phi (\varphi,\psi)\right\} \nonumber \\
&\, + \frac{1}{2} \left\{ \beta(\varphi,\psi) \partial_\mu \varphi \partial_\nu \varphi
+ \sigma(\varphi,\psi) \left( \partial_\mu \varphi \partial_\nu \psi
+ \partial_\nu \varphi \partial_\mu \psi \right)
+ \alpha (\varphi,\psi) \partial_\mu \psi \partial_\nu \psi \right\} \nonumber \\
&\, {\mathit- 2 \left( \nabla_\mu \nabla_\nu \zeta(\varphi,\psi)\right)R
+ 2 g_{\mu\nu} \left( \nabla^2 \zeta(\varphi,\psi)\right)R
+ 4 \left( \nabla_\rho \nabla_\mu \zeta(\varphi,\psi)\right)R_\nu^{\ \rho}
+ 4 \left( \nabla_\rho \nabla_\nu \zeta(\varphi,\psi)\right)R_\mu^{\ \rho}} \nonumber \\
&\, {\mathit- 4 \left( \nabla^2 \zeta(\varphi,\psi) \right)R_{\mu\nu}
 - 4g_{\mu\nu} \left( \nabla_\rho \nabla_\sigma \zeta(\varphi,\psi) \right) R^{\rho\sigma}
+ 4 \left(\nabla^\rho \nabla^\sigma \zeta(\varphi,\psi) \right) R_{\mu\rho\nu\sigma}
+ \frac{1}{2} T_{\mathrm{matter}\, \mu\nu}} \, ,
\end{align}
The   equation of motions of the scalar fields are obtained in the following expressions:
\begin{align}
\label{I100}
0 =& \frac{1}{2} \beta_\varphi \partial_\mu \varphi \partial^\mu \varphi
+ \beta\nabla^\mu \partial_\mu \varphi + \beta_\psi \partial_\mu \varphi \partial^\mu \psi
+ \left( \sigma_\psi - \frac{1}{2} \alpha_\varphi \right)\partial_\mu \psi \partial^\mu \psi
+ \sigma\nabla^\mu \partial_\mu \psi - \Phi_\varphi - \zeta_\varphi \mathcal{G} \, ,\nonumber \\
0 =& \left( - \frac{1}{2} \beta_\psi + \sigma_\varphi \right) \partial_\mu \varphi \partial^\mu \varphi
+ \sigma\nabla^\mu \partial_\mu \varphi
+ \frac{1}{2} \alpha_\psi \partial_\mu \psi \partial^\mu \psi + \alpha \nabla^\mu \partial_\mu \psi
+ \alpha_\varphi \partial_\mu \varphi \partial^\mu \psi - \Phi_\psi - \zeta_\psi \mathcal{G} \, ,
\end{align}
where $\beta_\varphi=\partial \beta(\varphi,\psi)/\partial \varphi$, etc. In
Eq.~(\ref{gb4bD44}), $\left(T_{\mathrm{matter}}\right)_{\mu\nu}$ represents  the energy-momentum tensor of matter.

Considering the following  spacetime, described by the metric\footnote{The justifications
that  metric~(\ref{GBiv1}) is a general one are presented in \cite{Nojiri:2020blr}.},
\begin{align}
\label{GBiv1}
ds^2 = - \e^{2\nu (t,r)} dt^2 + \e^{2\nu_1 (t,r)} dr^2 + r^2 \left( d\theta^2 + \sin^2\theta d\phi^2 \right)\, ,
\end{align}
where $\nu (t,r)$ and $\nu_1 (t,r)$ are two unknown function of time and radial coordinate. 
Furthermore, we use following identifications,
\begin{align}
\label{TSBH1}
\varphi=t\, , \qquad \qquad \psi=r\, .
\end{align}
  This assertion is not limiting in scope because the spherically symmetric solutions~(\ref{GBiv1}) of theory~(\ref{I88}) typically have $\varphi$ and $\psi$ varying with both $t$ and $r$. Once a solution is provided, the dependence of $\varphi$ and $\psi$ on $t$ and $r$ can be established, resulting in specific functions $\varphi(t,r)$ and $\psi(t,r)$ being identified.


For the purpose of eliminating any  ghosts that could appear we use the Lagrange
multiplier fields $\lambda_\varphi$ and $\lambda_\psi$,  that can
 be added  to the action $\left(S_{\mathrm{GR}}\right)_{\varphi\psi}$ given by Eq.~(\ref{I88}) as
$\left(S_{\mathrm{GR}}\right)_{\varphi\psi}\to \left(S_{\mathrm{GR}}\right)_{\varphi\psi} + S_\lambda$,
\begin{align}
\label{lambda1}
S_\lambda = \int d^4 x \sqrt{-g} \left[ \lambda_\varphi \left( \e^{-2\nu(t=\varphi, r=\psi)} \partial_\mu \varphi \partial^\mu \varphi + 1 \right)
+ \lambda_\psi \left( \e^{-2\lambda(t=\varphi, r=\psi)} \partial_\mu \psi \partial^\mu \psi - 1 \right) \right] \, .
\end{align}
Variations of Eq.~(\ref{lambda1}) with respect to $\lambda_\varphi$ and $\lambda_\psi$ yield,
\begin{align}
\label{lambda2}
0 = \e^{-2\nu(t=\varphi, r=\psi)} \partial_\mu \varphi \partial^\mu \varphi + 1 \, , \quad
0 = {\e^{-2\nu_1(t=\varphi, r=\psi)} \partial_\mu \psi \partial^\mu \psi - 1} \, ,
\end{align}
whose solutions are consistently given by Eq.~(\ref{TSBH1}).

In fact, by considering the perturbation from of Eq.~(\ref{TSBH1}) we get:
\begin{align}
\label{pert1}
\varphi=t + \delta \varphi \, , \qquad \psi=r + \delta \psi\, ,
\end{align}
By using Eq.~(\ref{lambda2}), we find
\begin{align}
\label{pert2}
\partial_t \left( { \e^{2\nu(t=\varphi, r=\psi)}} \delta \varphi \right) = \partial_r \left( { \e^{2\nu_1(t=\varphi, r=\psi)}} \delta \chi \right) = 0\, .
\end{align}
{ Eq.~(\ref{pert2}) tells that by imposing the initial condition $\delta\varphi=0$ (because $\varphi$ corresponds to the time coordinate)
and by imposing the boundary condition $\delta\psi\to 0$ when $r\to \infty$ (because $\psi$ corresponds to the radial coordinate),
we find that both of $\delta \varphi$ and $\delta \psi$ vanish in the whole spacetime, $\delta\varphi=0$ and $\delta\psi=0$
not only at the initial surface for $\delta\varphi$ and boundary surface for $\delta\psi$.
 Refs.~\cite{Nojiri:2023dvf, Nojiri:2023zlp, Elizalde:2023rds, Nojiri:2023ztz} tell that $\lambda_\varphi=\lambda_\psi=0$
consistently appear as a solution even in the model with the modified action $S + S_\lambda$.
This tells any solution of Eqs.~(\ref{gb4bD44}) and (\ref{I100}) which are based on the original action (\ref{I88}) is
a solution even for the modified model with the action $S + S_\lambda$.}

Upon modifying  $S_{\mathrm{GR} \varphi\psi} \to S_{\mathrm{GR} \varphi\psi} + S_\lambda$ in (\ref{lambda1}), the equations in (\ref{gb4bD44}) and (\ref{I100}) undergo the following modifications:
\begin{align}
\label{gb4bD4mod}
0= &\, \frac{1}{2\kappa^2}\left(- R_{\mu\nu} + \frac{1}{2} g_{\mu\nu} R\right) \nonumber \\
&\, + \frac{1}{2} g_{\mu\nu} \left\{
 - \frac{1}{2} \beta(\varphi,\psi) \partial_\rho \varphi \partial^\rho \varphi
 - \sigma(\varphi,\psi) \partial_\rho \varphi \partial^\rho \psi
 - \frac{1}{2} \alpha (\varphi,\psi) \partial_\rho \psi \partial^\rho \psi - \Phi (\varphi,\psi)\right\} \nonumber \\
&\, + \frac{1}{2} \left\{ \beta(\varphi,\psi) \partial_\mu \varphi \partial_\nu \varphi
+ \sigma(\varphi,\psi) \left( \partial_\mu \varphi \partial_\nu \psi
+ \partial_\nu \varphi \partial_\mu \psi \right)
+ \alpha (\varphi,\psi) \partial_\mu \psi \partial_\nu \psi \right\} \nonumber \\
&\, - 2 \left( \nabla_\mu \nabla_\nu \zeta(\varphi,\psi)\right)R
+ 2 g_{\mu\nu} \left( \nabla^2 \zeta(\varphi,\psi)\right)R
+ 4 \left( \nabla_\rho \nabla_\mu \zeta(\varphi,\psi)\right)R_\nu^{\ \rho}
+ 4 \left( \nabla_\rho \nabla_\nu \zeta(\varphi,\psi)\right)R_\mu^{\ \rho} \nonumber \\
&\, - 4 \left( \nabla^2 \zeta(\varphi,\psi) \right)R_{\mu\nu}
 - 4g_{\mu\nu} \left( \nabla_\rho \nabla_\sigma \zeta(\varphi,\psi) \right) R^{\rho\sigma}
+ 4 \left(\nabla^\rho \nabla^\sigma \zeta(\varphi,\psi) \right) R_{\mu\rho\nu\sigma} \nonumber \\
&\, + \frac{1}{2}g_{\mu\nu} \left\{ \lambda_\varphi \left( \e^{-2\nu(r=\psi)} \partial_\rho \varphi \partial^\rho \varphi + 1 \right)
+ \lambda_\psi \left( \e^{-2\lambda(r=\psi)} \partial_\rho \psi \partial^\rho \psi - 1 \right) \right\} \nonumber \\
&\, - \lambda_\varphi \e^{-2\nu(r=\psi)} \partial_\mu \varphi \partial_\nu \varphi
 - \lambda_\psi \e^{-2\lambda(r=\psi)} \partial_\mu \psi \partial_\nu \psi
+ \frac{1}{2} T_{\mathrm{matter}\, \mu\nu} \, , \\
\label{I10mod}
0 =& \frac{1}{2} \beta_\varphi \partial_\mu \varphi \partial^\mu \varphi
+ \beta\nabla^\mu \partial_\mu \varphi + \beta_\psi \partial_\mu \varphi \partial^\mu \psi
+ \left( \sigma_\psi - \frac{1}{2} \alpha_\varphi \right)\partial_\mu \psi \partial^\mu \psi
+ \sigma\nabla^\mu \partial_\mu \psi - \Phi_\varphi - \zeta_\varphi \mathcal{G} \nonumber \\
&\, - 2 \nabla^\mu \left( \lambda_\varphi \e^{-2\nu(r=\psi)} \partial_\mu \varphi \right) \, ,\nonumber \\
0 =& \left( - \frac{1}{2} \beta_\psi + \sigma_\varphi \right) \partial_\mu \varphi \partial^\mu \varphi
+ \sigma\nabla^\mu \partial_\mu \varphi
+ \frac{1}{2} \alpha_\psi \partial_\mu \psi \partial^\mu \psi + \alpha \nabla^\mu \partial_\mu \psi
+ \alpha_\varphi \partial_\mu \varphi \partial^\mu \psi - \Phi_\psi - \zeta_\psi \mathcal{G} \nonumber \\
&\, - 2 \nabla^\mu \left( \lambda_\psi \e^{-2\lambda(r=\psi)} \partial_\mu \psi \right)  \, .
\end{align}
Let's examine the solutions to the equations in (\ref{gb4bD44}) and (\ref{I100}) under the assumptions (\ref{GBiv1}) and (\ref{TSBH1}). Subsequently, the  $(t,t)$ and $(r,r)$ components in (\ref{gb4bD4mod}) yield:
\begin{align}
\label{lambdas}
\lambda_\varphi = \lambda_\psi=0\, .
\end{align}
The rest non-vanishing components  of Eq.~(\ref{gb4bD4mod}) are identically satisfied. Additionally, Eq.~(\ref{I10mod}) yields
\begin{align}
\label{lambdas2}
0 = \nabla^\mu \left( \lambda_\varphi \e^{-2\nu(r=\psi)} \partial_\mu \varphi \right)
= \nabla^\mu \left( \lambda_\psi \e^{-2\lambda(r=\psi)} \partial_\mu \psi \right) \, .
\end{align}
The satisfaction of Eq.~(\ref{lambdas2}) implies the satisfaction of Eq.~(\ref{lambdas}). Consequently, the solution to Eqs.~(\ref{gb4bD44}) and (\ref{I100}) corresponds to a solution of Eqs.~(\ref{gb4bD4mod}) and (\ref{I10mod}) associated with the modified action as
$S_{\mathrm{GR} \varphi\psi} \to S_{\mathrm{GR} \varphi\psi} + S_\lambda$ in (\ref{lambda1})
if $\lambda_\varphi$ and $\lambda_\psi$ are nil (\ref{lambdas}).
However, it's worth noting that for Eqs.~(\ref{gb4bD4mod}) and (\ref{I10mod}),  a solution exists where  $\lambda_\varphi$ and $\lambda_\psi$  do not equal zero.

The detailed perturbation of the field equations (\ref{gb4bD4mod}) is given in (see for example \cite{Nojiri:2023mbo,Elizalde:2023rds}. Therefore in this study we will not repeat these calculation but give the final perturbation form which yields:

\begin{align}
\label{gb4bD4BoneGW}
0=&\, \left[ \frac{1}{4\kappa^2} R + \frac{1}{2} \left\{
 - \frac{1}{2} \partial_\rho \varphi \partial^\rho \varphi - \Phi \right\}
 - 4 \left( \nabla_\rho \nabla_\sigma \zeta\right) R^{\rho\sigma} \right] h_{\mu\nu} \nonumber \\
&\, + \bigg[ \frac{1}{4} g_{\mu\nu} \left\{
 - \beta \partial^\tau \varphi \partial^\eta \varphi
 - \sigma \left(\partial^\tau \varphi \partial^\eta \psi+ \partial^\eta \varphi \partial^\tau \psi\right)
 - \alpha \partial^\tau \psi\partial^\eta \psi\right\} \nonumber \\
&\, - 2 g_{\mu\nu} \left( \nabla^\tau \nabla^\eta \zeta\right)R
 - 4 \left( \nabla^\tau \nabla_\mu \xi\right)R_\nu^{\ \eta} - 4 \left( \nabla^\tau \nabla_\nu \xi\right)R_\mu^{\ \eta}
+ 4 \left( \nabla^\tau \nabla^\eta \zeta\right)R_{\mu\nu} \nonumber \\
&\, + 4g_{\mu\nu} \left( \nabla^\tau \nabla_\sigma \zeta\right) R^{\eta\sigma}
+ 4g_{\mu\nu} \left( \nabla_{\rho} \nabla^\tau \zeta\right) R^{\rho\eta}
 - 4 \left(\nabla^\tau \nabla^\sigma \zeta\right) R_{\mu\ \, \nu\sigma}^{\ \, \eta}
 - 4 \left(\nabla^\rho \nabla^\tau \zeta\right) R_{\mu\rho\nu}^{\ \ \ \ \eta}
\bigg\} h_{\tau\eta} \nonumber \\
&\, + \frac{1}{2}\left\{ 2 \delta_\mu^{\ \eta} \delta_\nu^{\ \zeta} \left( \nabla_\kappa \zeta\right)R
 - 4 \delta_\rho^{\ \eta} \delta_\mu^{\ \zeta} \left( \nabla_\kappa \zeta\right)R_\nu^{\ \rho}
 - 4 \delta_\rho^{\ \eta} \delta_\nu^{\ \zeta} \left( \nabla_\kappa \zeta\right)R_\mu^{\ \rho} \right. \nonumber \\
&\, \left. + 4g_{\mu\nu} \delta_\rho^{\ \eta} \delta_\sigma^{\ \zeta} \left( \nabla_\kappa \zeta\right) R^{\rho\sigma}
 - 4 g^{\rho\eta} g^{\sigma\zeta} \left( \nabla_\kappa \zeta\right) R_{\mu\rho\nu\sigma}
\right\} g^{\kappa\lambda}\left( \nabla_\eta h_{\zeta\lambda} + \nabla_\zeta h_{\eta\lambda} - \nabla_\lambda h_{\eta\zeta} \right) \nonumber \\
&\, - \left\{ \frac{1}{4\kappa^2} g_{\mu\nu} - 2 \left( \nabla_\mu \nabla_\nu \zeta\right) + 2 g_{\mu\nu} \left( \nabla^2\zeta\right) \right\}
R^{\mu\nu} h_{\mu\nu} \nonumber \\
&\, + \frac{1}{2}\left\{ \left( - \frac{1}{2\kappa^2} - 4 \nabla^2 \zeta\right) \delta^\tau_{\ \mu} \delta^\eta_{\ \nu}
+ 4 \left( \nabla_\rho \nabla_\mu \zeta\right) \delta^\eta_{\ \nu} g^{\rho\tau}
+ 4 \left( \nabla_\rho \nabla_\nu \zeta\right) \delta^\tau_{\ \mu} g^{\rho\eta}
 - 4g_{\mu\nu} \nabla^\tau \nabla^\eta \zeta\right\} \nonumber \\
&\, \qquad \times \left\{ - \nabla^2 h_{\tau\eta} - 2R^{\lambda\ \varphi}_{\ \eta\ \tau}h_{\lambda\varphi}
+ R^\varphi_{\ \tau}h_{\varphi\eta} + R^\varphi_{\ \tau}h_{\varphi\eta} \right\} \nonumber \\
&\, + 2 \left(\nabla^\rho \nabla^\sigma \zeta\right)
\left\{ \nabla_\nu \nabla_\rho h_{\sigma\mu}
 - \nabla_\nu \nabla_\mu h_{\sigma\rho}
 - \nabla_\sigma \nabla_\rho h_{\nu\mu}
 + \nabla_\sigma \nabla_\mu h_{\nu\rho}
+ h_{\mu\varphi} R^\varphi_{\ \rho\nu\sigma}
 - h_{\rho\varphi} R^\varphi_{\ \mu\nu\sigma} \right\} \nonumber \\
&\, + \frac{1}{2}\frac{\partial T_{\mathrm{matter}\, \mu\nu}}{\partial g_{\tau\eta}}h_{\tau\eta} \, .
\end{align}
{ The observation of GW170817 gives the constraint on the propagating speed $c_\mathrm{GW}$ of the
gravitational wave as follows,
\begin{align}
\label{GWp9} \left| \frac{{c_\mathrm{GW}}^2}{c^2} - 1 \right| < 6
\times 10^{-15}\, ,
\end{align}
where $c$ denotes the speed of light.} To determine whether the speed at which the gravitational wave $h_{\mu\nu}$ propagates, denoted as $c_\mathrm{GW}$, is distinct from the speed of light $c$, we simply have to examine the components involving the second derivatives of $h_{\mu\nu}$.

\begin{align}
\label{second}
I_{\mu\nu} \equiv&\, I^{(1)}_{\mu\nu} + I^{(2)}_{\mu\nu} \, , \nonumber \\
I^{(1)}_{\mu\nu} \equiv&\, \frac{1}{2}\left\{ \left( - \frac{1}{2\kappa^2} - 4 \nabla^2 \zeta\right) \delta^\tau_{\ \mu} \delta^\eta_{\ \nu}
+ 4 \left( \nabla_\rho \nabla_\mu \xi\right) \delta^\eta_{\ \nu} g^{\rho\tau}
+ 4 \left( \nabla_\rho \nabla_\nu \xi\right) \delta^\tau_{\ \mu} g^{\rho\eta}
 - 4g_{\mu\nu} \nabla^\tau \nabla^\eta \zeta\right\} \nabla^2 h_{\tau\eta} \, , \nonumber \\
I^{(2)}_{\mu\nu} \equiv &\, 2 \left(\nabla^\rho \nabla^\sigma \zeta\right)
\left\{ \nabla_\nu \nabla_\rho h_{\sigma\mu}
 - \nabla_\nu \nabla_\mu h_{\sigma\rho}
 - \nabla_\sigma \nabla_\rho h_{\nu\mu}
 + \nabla_\sigma \nabla_\mu h_{\nu\rho} \right\} \, .
\end{align}

Now we are ready to derive time dependent black hole in the context of EGB gravity coupled with  $\varphi$ and $\psi$.
\section{Time dependent black hole}\label{sec2}
Now let us apply the line element given by Eq. (\ref{GBiv1}) to the field equations (\ref{gb4bD44}) and get:
\begin{align}
&t\,t-component:\\
&{\mathrm {\frac {1-2\nu'_1 \left( t,r \right)\,r-  {e^{2\nu_1 \left( t,r \right) }}}{{r}^{2} {e^{2\nu_1 \left( t,r \right) } }  }}=\frac{-{\kappa}^{2}}{2  {r} ^{2} {e^{2[\nu \left( t,r \right)+2\nu_1 \left( t,r \right)] }}} \left[ 16{e^{ 2\nu \left( t,r \right) }} \left\{ 1- e^{2\nu_1 \left( t,r \right)}   \right\}\xi'' \left( t,r \right) +16  \zeta' \left( t,r \right){ e^{2\nu \left( t,r \right) }}  \left\{{ e^{2\nu_1 \left( t,r \right) }} -3 \right\}\nu'_1 \left( t,r \right)  \right.}\nonumber\\
 &\left.{\mathrm+ \left\{ 16 \left({e^{2\nu_1 \left( t,r \right) }} -1\right)   \dot\zeta \left( t,r \right)  \dot\nu_1 \left( t,r \right) + \left[  \left\{ 2{e^{2\nu \left( t,r \right) }}\Phi \left( t,r \right) +\beta \left( t,r \right)   \dot\varphi^2 \left( t \right)  \right\}  { e^{2\nu_1 \left( t,r \right) }}+ {e^{2 \nu \left( t,r \right) }}\psi'^2 \left( r \right)\alpha \left( t,r \right)  \right] {r}^{2} \right\}  {e^{2\nu_1 \left( t,r \right) }} } \right],\nonumber\\
 &t\,r-component:\\
 & {\mathrm {\frac {2\dot\nu_1 \left( t,r \right) } { r{e^{2\nu_1 \left( t,r \right) }} }}= \frac {{\kappa}^{2}}{{r}^{2}  {e^{4\nu_1 \left( t,r \right) }} } \left[  \left\{ 8-8  {e^{2\nu_1 \left( t,r \right) }}  \right\} \dot\zeta' \left( t,r \right) + \left\{8  { e^{2\nu_1 \left( t,r \right) }}-24 \right\} \zeta' \left( t,r \right) \dot\nu_1 \left( t,r \right) + \left\{ 8  {e^{2\nu_1 \left( t,r \right) }}  -8\right\}   \dot\zeta \left( t,r \right) \nu' \left( t,r \right)\right.}\nonumber\\
 &\left. {\mathit+\gamma \left( t,r \right) \dot\varphi \left( t \right)   \psi' \left( r \right)  {r}^{2} {e^{2\nu_1 \left( t,r \right) }}} \right]\,,\nonumber\\
 &r\,t-component=e^{2\nu_1 \left( t,r \right) }(t\,r-component):\\
 &r\,r-component:\\
&{\mathrm {\frac {1+2\nu' \left( t,r \right)\, r-  {e^{2\nu_1 \left( t,r \right) }}}{{r}^{2} {e^{2\nu_1 \left( t,r \right) } }  }}=\frac{{\kappa}^{2}}{2 {r} ^{2} {e^{2[\nu \left( t,r \right)+2\nu_1 \left( t,r \right) ]}}} \left[ 16{e^{ 2\nu_1 \left( t,r \right) }} \left\{ 1- e^{2\nu_1 \left( t,r \right)}   \right\}\ddot\xi \left( t,r \right) +16  \zeta' \left( t,r \right){ e^{2\nu \left( t,r \right) }}  \left\{{ e^{2\nu_1 \left( t,r \right) }} -3 \right\}\nu'_1 \left( t,r \right)  \right.}\nonumber\\
 &\left.{\mathrm+ \left\{ 16 \left({e^{2\nu_1 \left( t,r \right) }} -1\right)   \dot\zeta \left( t,r \right)  \dot\nu_1 \left( t,r \right) + \left[\left\{ \beta \left( t,r \right)   \dot\varphi^2 \left( t \right) -2{e^{2\nu \left( t,r \right) }}\Phi \left( t,r \right)  \right\}  { e^{2\nu_1 \left( t,r \right) }}+ {e^{2 \nu \left( t,r \right) }}\psi'^2 \left( r \right)\alpha \left( t,r \right)  \right] {r}^{2} \right\}  {e^{2\nu_1 \left( t,r \right) }} }\right],\nonumber\\
 &\theta\,\theta=\phi\,\phi-component:\\
 &{\mathrm{\frac {\left[(1+\nu' \left( t,r \right)\, r)(\nu' \left( t,r \right)-\nu'_1 \left( t,r \right))+\nu'' \left( t,r \right)\, r\right]{e^{-2\nu_1 \left( t,r \right)}}+r\left[\dot\nu_1 \left( t,r \right)\dot\nu \left( t,r \right)-\ddot\nu_1 \left( t,r \right)-\dot\nu^2_1 \left( t,r \right)\right]{e^{-2\nu \left( t,r \right)}}}{{r} }}}\nonumber\\
 &{\mathrm=\frac{\kappa^2}{2r}\left[16\left\{\ddot\nu_1 \left( t,r \right) \zeta'\left( t,r \right)+2\dot\nu_1\left( t,r \right)\dot\zeta'\left( t,r \right)-\nu'_1\left( t,r \right) \ddot\zeta\left( t,r \right) -\dot\nu_1\left( t,r \right)[\dot\nu_1\left( t,r \right)+\dot\nu\left( t,r \right)]\right\}\zeta'\left( t,r \right)\right.}\nonumber\\
 &\left.{\mathrm-16\dot\zeta\left( t,r \right)\left\{\dot\nu_1\left( t,r \right)\nu'\left( t,r \right)-\nu'_1\left( t,r \right)\dot\nu\left( t,r \right)\right\}{e^{-2[\nu_1 \left( t,r \right)+\nu \left( t,r \right)] }}-{e^{-4\nu_1 \left( t,r \right)}}\left\{16\zeta'\left( t,r \right)\nu''\left( t,r \right)+16\nu'\left( t,r \right)\right.}\right.\nonumber\\
 &\left.\left.{\mathrm\times[\zeta''+\zeta'\{\nu'\left( t,r \right)-3\nu'_1\left( t,r \right)\}]}\right\}+{\mathit r\left\{{e^{-2\nu \left( t,r \right)}}\beta\left( t,r \right)\dot\varphi^2(r)-{e^{-2\nu_1 \left( t,r \right)}}\alpha\left( t,r \right)\psi'^2(r)-2\Phi\left( t,r \right)\right\}}\right]
\end{align}
The above system of differential equations can be solved for the coefficients of the scalar fields in addition to the potential to have the form:
\begin{align}\label{scal}
&{\mathrm \beta(t,r)=\frac { {e^{-4\,\nu_1 \left( t,r \right) }}}{{ \kappa}^{2}{r}^{2}} \left[  \left\{ 8\,{\kappa}^{2} \zeta'' \left( t,r \right) + \nu'' \left( t,r \right)  {r}^{2}-8\,{ \kappa}^{2}\zeta' \left( t,r \right)  \nu_1' \left( t,r \right) + \left(r -{r}^{2}\nu' \left( t, r \right) \right) \nu_1' \left( t,r \right) +r\nu' \left( t,r \right) -1+ \nu'^2 \left( t,r \right){r}^{2} \right\}\right.}\nonumber\\
 &\left.{\mathrm {e^{2\nu ( t,r ) +2 \nu_1 ( t,r ) }}+e^{2\nu ( t,r ) +4 \nu_1 \left( t,r \right) }+ \left[8{\kappa}^{2} \ddot\zeta \left( t,r \right) \nu'_1 \left( t,r \right) r -8{\kappa}^{2} \zeta' \left( t,r \right) \ddot\nu_1 \left( t,r \right) r-16{\kappa}^{2}\dot\nu_1 \left( t,r \right) \dot\zeta' \left( t,r \right)   r+8{\kappa}^{2}\dot\nu_1^2 \left( t,r \right) \zeta' \left( t,r \right)  r\right.}\right.\nonumber\\
 &\left.\left.{\mathrm+ \left( 8{ \kappa}^{2} \zeta' \left( t,r \right) \dot\nu \left( t, r \right) r+8{\kappa}^{2}\dot\zeta \left( t,r \right) +8{\kappa}^{2} \dot\zeta \left( t,r \right)\nu' \left( t,r \right) r \right) \dot\nu_1 \left( t,r \right)} -8{\kappa}^{2}\dot\zeta \left( t,r \right)  \dot\nu \left( t,r \right)\nu_1' \left( t,r \right) r \right] {e^{2\,\nu_1 \left( t,r \right) }}+ \left[  \left({\mathit 8{\kappa}^{2}\nu' \left( t,r \right) r}\right.\right.\right.\nonumber\\
 &\left.\left.\left.-{\mathrm 8\,{\kappa}^{2}} \right) {\mathit \zeta'' \left( t,r \right) +8\,{ \kappa}^{2} \zeta' \left( t,r \right) \nu'' \left( t,r \right)   r+ \left(  \left( -24\,{\kappa}^{2} \nu' \left( t,r \right)r+24\,{\kappa}^{2} \right)  \nu_1' \left( t,r \right) +8\,{\kappa}^{2} \nu' \left( t,r \right)^{2}r \right) \zeta' \left( t,r \right)}  \right]{\mathit {e^{2\, \nu \left( t,r \right) }}}\right.\nonumber\\
 &\left.{\mathrm+ \left[ \left(  \dot\nu \left( t,r \right) {r}^{2}-8\,{\kappa}^{2}\dot\zeta \left( t,r \right)  \right) \dot\nu_1 \left( t,r \right)  - \ddot\nu_1^2 \left( t,r \right) - \dot\nu_1^2 \left( t,r \right)  {r}^{2} \right] {e^{4\,\nu_1 \left( t,r \right) }}} \right]
\,,\nonumber\\
&{\mathrm \gamma(t,r)={\frac {2{e^{-2\,\nu_1 \left( t,r \right) }}}{{ \kappa}^{2}{r}^{2}}} \left[  \left\{ 4\,{\kappa}^{2}\dot\zeta' \left( t,r \right) + \left( r-4\,{\kappa}^{2}\zeta' \left( t,r \right)  \right) \dot\nu_1 \left( t,r \right) -4\,{\kappa}^{2} \dot\zeta \left( t,r \right) \nu' \left( t,r \right)  \right\}{e^{2\, \nu_1 \left( t,r \right) }}+4\,{\kappa}^{2} \left( 3\, \zeta' \left( t,r \right)  \dot\nu_1 \left( t,r \right)\right.\right.}\nonumber\\
 &\left.\left.{\mathrm -\dot\zeta' \left( t,r \right) + \dot\zeta \left( t,r \right)  \nu' \left( t,r \right)}  \right)  \right] \,,\nonumber\\
 &{\mathit\alpha(t,r)=-\frac {{e^{-2\nu \left( t,r \right) -2\nu_1 \left( t,r \right) }}}{{\kappa}^{2}{r}^{2}} \left[  \left\{  \nu'' \left( t,r \right) {r}^{2}+\nu'^2 \left( t,r \right)  {r}^{2}+ \left( 8{\kappa}^{2}\zeta' \left( t, r \right) -r- \nu'^2_1 \left( t,r \right)   \right) \nu' \left( t,r \right) - \nu'_1 \left( t,r \right)  r-1 \right\} {e^{2\nu \left( t,r \right) +2\nu_1 \left( t,r \right) }}\right.}\nonumber\\
 &\left.{\mathrm+{e^{2\nu \left( t,r \right) +4\nu_1 \left( t,r \right) }}+ \left[  \left( 8{\kappa} ^{2} \nu'_1 \left( t,r \right)r+8{\kappa}^{2} \right) \ddot\zeta \left( t,r \right) -8{\kappa}^{2} \zeta' \left( t,r \right) \ddot\nu_1 \left( t,r \right)  r -16{\kappa}^{2} \dot\nu_1 \left( t,r \right) \dot\zeta' \left( t,r \right)  r+8{\kappa}^{2} \dot\zeta \left( t,r \right)  \dot\nu_1 \left( t,r \right) \nu' \left( t,r \right) r\right.}\right.\nonumber\\
 &\left.\left.{\mathrm+ 8\left( {\kappa}^{2}\dot\nu \left( t,r \right) \dot\nu_1 \left( t,r \right)  r+{\kappa}^{2} \dot\nu^2_1 \left( t,r \right) r \right) \zeta' \left( t,r \right) -8{\kappa}^{2} \dot\zeta \left( t,r \right) \dot\nu \left( t,r \right) -8{\kappa}^{2} \dot\zeta \left( t,r \right)  \dot\nu \left( t,r \right) \nu'_1 \left( t,r \right)   r} \right] {e^{ 2\nu_1 \left( t,r \right) }}+ \left[ \dot\nu \left( t,r \right)\dot\nu_1 \left( t,r \right) {r}^{2} \right.\right.\nonumber\\
 &\left.\left.{\mathrm- \ddot\nu_1 \left( t,r \right)   {r}^{2}- \dot\nu^2_1 \left( t,r \right){r}^{2}+8{\kappa}^{2} \dot\zeta \left( t,r \right)\dot\nu \left( t,r \right) -8{ \kappa}^{2}\ddot\zeta \left( t,r \right) } \right] {e^{4\nu_1 \left( t,r \right) }}+ \left[ 8 {\kappa}^{2} \zeta' \left( t,r \right) \nu'' \left( t,r \right)  r+8{\kappa}^{2} \zeta'' \left( t,r \right) \nu' \left( t,r \right)  r\right.\right.\nonumber\\
 &\left.\left.{\mathrm+8{\kappa}^{2} \zeta' \left( t,r \right) \nu'^2 \left( t,r \right)r-24 \left( {\kappa}^{2} \nu'_1 \left( t,r \right) r+{\kappa}^{2} \right) \zeta' \left( t,r \right)  \nu' \left( t,r \right) } \right] {e^{2\nu \left( t, r \right) }} \right]\,,\nonumber\\
 &{\mathrm\Phi(t,r)=-\frac {{e^{-2\,\nu \left( t,r \right) -4\,\nu_1 \left( t,r \right) }}}{{\kappa}^{2}{r}^{2}} \left[ \left\{ -4\,{\kappa}^{2}\zeta'' \left( t,r \right) -4\,{\kappa}^{2} \left[ \nu' \left( t,r \right) -\nu'_1 \left( t,r \right)  \right] \zeta' \left( t,r \right) - \nu'_1 \left( t,r \right) r+1+r\nu' \left( t,r \right)  \right\} {e ^{2\,\nu \left( t,r \right) +2\,\nu_1 \left( t,r \right) }}\right.}\nonumber\\
 &\left.{\mathrm-{e ^{2\,\nu \left( t,r \right) +4\,\nu_1 \left( t,r \right) }}+ \left\{ -4\,{\kappa}^{2}\ddot\zeta \left( t, r \right) + \left( -4\,{\kappa}^{2}\dot \nu_1 \left( t,r \right) +4 \dot \nu \left( t,r \right)  {\kappa}^{2} \right) \dot\zeta \left( t,r \right)  \right\} {e^{2\,\nu_1 \left( t,r \right) }}+ \left\{ 4\,{\kappa}^{2}\ddot\zeta \left( t,r \right) + \left( -4\, \dot\nu \left( t,r \right){\kappa}^{2}\right.\right.}\right.\nonumber\\
 &\left.\left.\left.{\mathrm+4 \,{\kappa}^{2}\dot\nu_1 \left( t,r \right)}  \right) \dot\zeta \left( t,r \right)  \right\} {e^{4\,\nu_1 \left( t,r \right) }}+ \left\{ 4 \,{\kappa}^{2}\zeta'' \left( t,r \right) + \left\{ 12\,{\kappa}^{2}\nu' \left( t,r \right) -12\,{\kappa}^{2}\nu'_1 \left( t,r \right)  \right\} \zeta' \left( t,r \right)  \right\} {e^{2\,\nu \left( t,r \right) }} \right]\,.
\end{align}
\section{Time dependent black hole solutions}\label{sec3}
In this section, we are going to derive the coefficients of the  scalar fields, $\varphi$ and $\psi$, that are  time dependent black hole.
\subsection{First time dependent black hole}
In this case we take the ansatzs of the BH to have the from\footnote{{ In this study, we assume  $\nu(t,r)$ is equal to $-\nu_1(t,r)$, as seen in the Schwarzschild. The line element in spherical spacetime is expressed using the areal radius $r$ with $g_{tt}g_{rr}=-1$, where $r$ serves as an affine parameter along radial null geodesics \cite{Jacobson:2007tj}. Additionally, this spacetime possesses unique algebraic characteristics: the Ricci tensor's double projection onto radial null vectors results in zero \cite{Jacobson:2007tj,BondiKilmister,Dadhich:2015wgx}. On the other hand, the Ricci tensor's restriction to the $\left(t, r\right)$ submanifold is directly related to the restriction of the metric $g_{\mu\nu}$ to this specific subspace \cite{Jacobson:2007tj}.  Several solutions in general relativity or alternative gravities are characterized by the condition $g_{tt}g_{rr}=-1$, such as vacuum solutions, electrovacuum solutions with Maxwell or nonlinear Born-Infeld electrodynamics, and the string hedgehog global monopole \cite{Jacobson:2007tj}, also in higher dimensions, as shown in  \cite{Barriola:1989hx,Guendelman:1991qb}}.}:
\begin{align}\label{ans1}
e^{2\nu(t,r)}=e^{-2\nu_1(t,r)}=1-\frac{2M(t)}{r}\,,
\end{align}
where the Misner-Sharp-Hernandez mass $M(t)$ is positive and depends only on time. { A negative mass, $M$, indicates a violation of the energy conditions and makes apparent horizons impossible.}
{ The Ricci scalar related to ansatz (\ref{ans1}) is given by:
\begin{equation} \label{TSBHFR9} R= \left( 1 - \frac{2M}{r} \right)^{-1}
\left[ \frac{2 \ddot M}{r}
+ \frac{\left( \frac{2 \dot M}{r} \right)^2}{1 - \frac{2 M}{r}} \right]\,.
  \end{equation} Equation (\ref{TSBHFR9}) demonstrates that when $M$ remains constant, $R$ becomes zero, indicating the geometry is the same as Schwarzschild geometry. In this scenario, there is a single visible horizon, with a surface radius of $r(t)=2M(t)$.  This horizon, which is constantly changing, represents a black hole horizon as it is a unique solution to the equation $\nabla^\alpha r \nabla_\alpha r=0.$ A curvature singularity occurs when $2M=r$ given that $\frac{2 \ddot M}{r} \left( 1 - \frac{2 M}{r} \right)$. The value of $\left( \frac{2 \dot M}{r} \right)^2$ decreases significantly as $\mathcal{O} \left( \left( 1 - \frac{2 M}{r} \right)^2 \right)$ approaches zero when $2 M=r$, which is not possible.}
  Using Eq. (\ref{ans1}) in Eqs.~ (\ref{scal}) we present the coefficient of the scalar fields and the potential in the Appendix due to its lengthy:

\subsection{Second time dependent black hole}
The second time dependent ansatz of the BH  that we will deal takes the form:
\begin{align}\label{ans2}
e^{2\nu(t,r)}=e^{-2\nu_1(t,r)}=\left(1-\frac{r_0}{r}\right)\frac{t_0}{t}\,,
\end{align}
 with $r_0$ and $t_0$ being positive constants. The Misner-Sharp-Hernandez mass $M(t)$ of the above line element takes the form:
\begin{align}\label{mass2}
M(t)=\frac{r t-t_0 r+t_0r_0}{2t}\,.
\end{align}
 Using Eq. (\ref{ans2}) in Eqs.~(\ref{scal}) we get the coefficient of the scalar fields   and the potential in the Appendix.
{ Equation (\ref{mass2}) demonstrates that only a single apparent horizon exists at $r =2M\left(t,r\right)$, resulting in $r=r_0$. Since this is a single root, we have a black hole apparent
horizon. Equation (\ref{mass2}) reveals that there is just a single visible horizon situated at $r=2M\left(t,r\right)$, where $r=r_0$.  This horizon is both a null surface and an event horizon.
{ The Ricci curvature of Eq.(\ref{mass2}) has the form: \begin{equation} \label{Ex2c} R= \frac{2}{r^2}
\left( 1 - \frac{t_0}{t} \right) \,. \end{equation} The  geometry of Eq.~(\ref{Ex2c}) has no
spacetime singularities except for the usual one at $r=0$ and the Big Bang at $t=0$.}}
\subsection{Third time dependent black hole}
The second time-dependent ansatz of the BH  that we will deal with takes the form:
\begin{align}\label{ans3}
e^{2\nu(t,r)}=e^{-2\nu_1(t,r)}=\frac{1-\frac{r_0}{r}}{1+\frac{tr_0}{rt_0}}\,,
\end{align}
where the Misner-Sharp-Hernandez mass $M(t)$ of the above line element take the form:
\begin{align}\label{mass3}
M(t)=\frac{rr_0(t_0+t)}{2(rt_0+tr_0)}\,.
\end{align}
{ Equation (\ref{mass3}) displays that the apparent horizons can be found at $r=2M$, resulting in the sole root being $r=r_0$. This represents a lone root and the circumference of a stationary black hole event horizon.  The mass located at this boundary is equal to half the radius $M(r_0)= r_0/2$ and remains constant over time. As $r$ approaches infinity, the relationship $\e^{2\nu} \, = \, \e^{-2\nu_1} \to 1$ indicates that the geometry is flat at infinity.  As $t$ approaches infinity while $r$ remains constant, the metric can be approximated by $\e^{2\nu} \simeq \frac{t_0 \, r}{t r_0} \left(1 - \frac{r_0}{r}\right)$. Next, by defining a new time parameter $\tau$ as $d\tau =\frac{dt}{\sqrt{t}} $ (or $\tau (t)= 2 \sqrt{t} \,$), the metric equation can be rewritten as:
\begin{align} \label{Ex3A2} ds^2 = - \frac{t_0\, r}{r_0} \left(1 - \frac{r_0}{r}\right) d\tau^2
+ \frac{r_0 \tau^2 }{4 t_0 \, r \left(1 - r_0/r \right)} \, dr^2
  + r^2 \left( d\vartheta^2 + \sin^2\vartheta \, d\varphi^2 \right)\,,
    \end{align} as the time $\tau$ (or $t$) increases, the radial direction expands resembling a throat. The horizontal radius of the horizon stays the same. The factor depending on time $\left(\frac{r_0}{4t_0 \, r \left(1 - \frac{r_0}{r}\right)} \, \tau^2\right)$ affects only $dr^2$ and not the angular component of the metric, unlike in a FLRW universe with a central object (the metric approaches flat instead of FLRW at infinity) \cite{Vbook}. { The Ricci scalar of the geometry~(\ref{ans3}) is \begin{equation}
\label{ex3a6} R = \frac{2 \left(1 +
\frac{t_0}{t}\right)\frac{r_0}{r^3}}{\left(1
+ \frac{t r_0}{t_0 \, r}\right)^3} \left[ 1 - \frac{t}{t_0} + 2 \left( 1 -
  \frac{t}{t_0} \right) \frac{t r_0}{t_0 \, r} + \frac{t^2 r_0^2}{t_0^2
\, r^2 }
  \right] \, , \end{equation} which is continuous at the horizon $r=r_0$, despite the presence of singularities at $r=0, t=0$, and $t\to \infty$.}}
 Using Eq. (\ref{ans2}) in Eq.~(\ref{scal})  we present the coefficient of the scalar fields as well as the potential in the Appendix.
\section{Gravitational wave propagation}\label{sec4}
Now, let us investigate the  GW propagation. The EGB theory requires that the propagation speed of gravitational waves with one scalar field corresponds to the speed of light in the FRW   spacetime has been studied in \cite{Oikonomou:2020sij}.
The case of propagation of gravitational waves for a spacetime with two scalar fields was provided in \cite{Nojiri:2023jtf}.
Now let us use the condition given by Eq.~(\ref{second}).  In our examination of the GW propagation velocity, denoted as $c_\mathrm{GW}$,  our focus is solely on expressions that encompass the $2^{nd.}$ order differentiation  of the GW tensor $h_{\mu\nu}$.  Given the assumption of minimal interaction between gravity and matter, the contribution from the field of matter does not interact through the differentiation of the gravitational wave tensor $h_{\mu\nu}$. As a result, they are not present in the tensor $I_{\mu\nu}$. Put simply, the existence of matter has no impact on the GW's propagation velocity. Importantly, the tensor $I^{(1)}_{\mu\nu}$ does not change the GW's speed, which continues to match that of light. In general, the tensor $I^{(2)}_{\mu\nu}$ has the potential to change the GW's speed, deviating it from the speed of light. In this analysis, we have omitted the disturbances of the scalar fields $\varphi$ and $\psi$, as these fluctuations can be rendered nonexistent by the constraints outlined in Eq.~(\ref{lambda2}). When there are no constraints present, or if only a single scalar field interacts with the Gauss-Bonnet invariant, the scalar mode perturbations become intertwined with the scalar modes within the metric's fluctuation. Provided that we focus on the massless spin-two modes, representative of the conventional GW, these modes become independent from the scalar mode at the foremost order. When taking into account the second-order perturbation, the quadratic moment can emerge as a result of the scalar mode's perturbation. This quadratic moment acts as the origin of the GW. Hence, the propagation characteristics of the GW may vary between scenarios with the constraints, as specified in Eq.~(\ref{lambda2}), and those without them.

In the context of the metric outlined in Eq.~(\ref{GBiv1}), the sole connection coefficients that are not zero are the following:
\begin{align}\label{GBv}
&\left\{_{t}^{t}{}_{ t}\right\}=\dot\mu \, , \qquad \left\{_{t}^{r}{}_{ t}\right\}
= \e^{-2(\nu_1 - \nu)}\nu' \, , \qquad \left\{_{t}^{t}{}_{ r}\right\}=\left\{_{r}^{t}{}_{ t}\right\}=\nu'\,
, \qquad \left\{_{r}^{t}{}_{ r}\right\}= \e^{2\nu_1 - 2\mu}\dot\nu_1\, , \qquad
\left\{_{t}^{r}{}_{ r}\right\}= \left\{_{r}^{r}{}_{ t}\right\} = \dot\nu_1 \, , \nonumber\\
& \left\{_{r}^{r}{}_{ r}\right\}=\nu_1'\, , \quad \left\{_{j}^{i}{}_{ k}\right\}=\bar{\left\{_{t}^{t}{}_{ t}\right\}}\, ,\qquad
\left\{_{i}^{r}{}_{ j}\right\} -\e^{-2\nu_1}r \bar{g}_{ij} \, ,\qquad  \left\{_{r}^{i}{}_{ j}\right\}=\left\{_{j}^{i}{}_{ r}\right\}\frac{1}{r} \, \delta^i_{\ j}\,,
\end{align} with $\bar{\left\{_{j}^{i}{}_{ k}\right\}}$  represents the metric connection of $\bar{g}_{ij}$, an overdot signifies differentiation with respect to time $t$, and a prime indicates differentiation with respect to radial distance  $r$.
Using Eq. (\ref{GBv}) we get the following  formula for the Riemann tensor:
\begin{align}
\label{curvatures}
R_{rtrt} = & - \e^{2\nu_1} \left\{ \ddot\nu_1 + \left( \dot\nu_1 - \dot\nu \right) \dot\nu_1 \right\}
+ \e^{2\nu}\left\{ \nu'' + \left(\nu' - \nu_1'\right)\nu' \right\} \, ,\nonumber \\
R_{titj} =& \, r\nu' \e^{-2\nu_1 + 2\nu} \bar{g}_{ij} \, ,\nonumber \\
R_{rirj} =& \, \nu_1' r \bar{ g}_{ij} \, ,\quad {R_{tirj}= \dot\nu_1 r \bar{ g}_{ij} } \, , \quad
R_{ijkl} = \left( 1 - \e^{-2\nu_1}\right) r^2 \left(\bar{g}_{ik} \bar{g}_{jl} - \bar{g}_{il} \bar{g}_{jk} \right)\, ,\nonumber \\
R_{tt} =& - \left\{ \ddot\nu_1 + \left( \dot\nu_1 - \dot\nu \right) \dot\nu_1 \right\}
+ \e^{-2\nu_1 + 2\nu} \left\{ \nu'' + \left(\nu' - \nu_1'\right)\nu' + \frac{2\nu'}{r}\right\} \, ,\nonumber \\
R_{rr} =& \, \e^{2\nu_1 - 2\nu} \left\{ \ddot\nu_1 + \left( \dot\nu_1 - \dot\nu \right) \dot\nu_1 \right\}
 - \left\{ \nu'' + \left(\nu' - \nu_1'\right)\nu' \right\}
+ \frac{2 \nu_1'}{r} \, ,\quad
R_{tr} =R_{rt} = \frac{2\dot\nu_1}{r} \, , \nonumber \\
R_{ij} =&\, \left\{ 1 + \left\{ - 1 - r \left(\nu' - \nu_1' \right)\right\} \e^{-2\nu_1}\right\} \bar{g}_{ij}\ , \nonumber \\
R=& \, 2 \e^{-2 \nu} \left\{ \ddot\nu_1 + \left( \dot\nu_1 -
\dot\nu \right) \dot\nu_1 \right\} + \e^{-2\nu_1}\left\{ -
2\nu'' - 2\left(\nu' - \nu_1'\right)\nu' - \frac{4\left(\nu' -
\nu_1'\right)}{r} + \frac{2\e^{2\nu_1} - 2}{r^2} \right\} \, ,
\end{align}
and
\begin{align}
\label{xis}
\nabla_t \nabla_t \zeta =&\, \ddot\zeta - \dot\nu \dot\zeta - \e^{-2(\nu_1 - \nu)}\nu' \zeta' \, , \quad
\nabla_r \nabla_r \zeta = \zeta'' - \e^{2\nu_1 - 2\nu} \dot\nu_1 \dot\zeta - \nu_1' \zeta' \, , \quad
\nabla_t \nabla_r \zeta = \nabla_r \nabla_t \zeta = {\dot\zeta}' - \nu' \dot\zeta - \dot\nu_1 \zeta' \, , \nonumber \\
\nabla_i \nabla_j \zeta =&\, - \e^{-2\nu_1} r {\bar g}_{ij} \zeta'
\, , \quad \nabla^2 \zeta = - \e^{-2\nu} \left( \ddot\zeta - \left(
\dot\nu - \dot\nu_1\right) \dot\zeta \right) + \e^{-2\nu_1}
\left( \zeta'' + \left( \nu' - \nu_1' - \frac{2}{r} \right) \zeta'
\right) \, ,
\end{align}
The metric tensor $(\bar{g}_{ij})$ can be expressed as
$$\sum_{i,j=1}^2\bar{g}_{ij} dx^i dx^j = d\vartheta^2 + \sin^2\vartheta , d\varphi^2\,,$$
where $\left(x^1=\vartheta, x^2=\varphi\right)$, and
$\bar{\left\{_{j}^{i}{}_{ k}\right\}}$   are the coordinates, denoted by $x^1=\vartheta$ and $x^2=\varphi$, respectively. Additionally, $\bar{\left\{_{j}^{i}{}_{ k}\right\}}$   denotes the connection coefficients associated with the metric tensor $\bar{g}_{ij}$.
The symbols ``dot'' and ``prime'' signify the derivatives w.r.t. $r$ and $t$, respectively. We will proceed to analyze the  GW that is propagating outwards along the radial axis that has the form:
\begin{align}
\label{hij} h_{ij} = \frac{\mathrm{Re} \left( \e^{-i\varpi t + i k r} \right) h_{ij}^{(0)}}{r}
\quad \left( i,j =\theta, \phi, \quad
\sum_i h_{\ \ i}^{(0)\ i}=0 \right)\,  .
\end{align}
We proceed with the assumption that $k$ is sufficiently large, which leads us to retain terms that are quadratic in relation to $k $ and/or $\varpi$, while disregarding any terms that are linear with respect to $k$ or $\varpi$.
Therefore,  (\ref{second}) and  (\ref{xis}) yields:
\begin{align}
\label{dis1}
0=&\, \left[ - \frac{1}{4\kappa^2} - 2 \left\{ - \e^{-2\nu} \left( 2 \ddot\zeta - \left( 2\dot\nu - \dot\nu_1\right) \dot\zeta \right) + \e^{-2\nu_1}
\left( \zeta'' + \left( 2 \nu' - \nu_1' \right) \zeta' \right) \right\} \right] \e^{-2\nu} \varpi^2 \nonumber \\
&\, - \left[ - \frac{1}{4\kappa^2} - 2 \left\{ - \e^{-2\nu} \left( \ddot\zeta - \left( \dot\nu - 2 \dot\nu_1\right) \dot\zeta \right) + \e^{-2\nu_1}
\left( 2 \zeta'' + \left( \nu' - 2 \nu_1' \right) \zeta' \right) \right\} \right] \e^{-2\nu_1} k^2 \nonumber \\
&\, - 4 \left( {\dot\zeta}' - \nu' \dot\zeta - \dot\nu_1 \zeta' \right) \e^{-2\nu - 2\nu_1} k\varpi \, .
\end{align}
Upon assuming that $\zeta$ is sufficiently small, it is observed that,
\begin{align}
\label{dis2}
\frac{\omega}{k} = \e^{- \nu_1 +\nu } &\, \left[\pm\left\{ 1-4 \kappa^2 \left( -\ddot\zeta + \dot\zeta \left( \dot \nu_1 +\dot\nu \right) \right) \e^{ -2\nu}
 -4 \left\{ -\zeta'' + \zeta' \left(\nu_1' +\nu' \right) \right\} \kappa^2 \e^{-2 \nu_1} \right\} - 8 \e^{-\nu_1-\nu}\kappa^2 \left( \dot\zeta \nu' + \zeta' \dot\nu_1 -\dot\zeta' \right)\right] \,.
\end{align}
Therefore, $c$ is determined by the equation $c = e^{-\nu_1 + \nu}$. Consequently, when
\[
\left\{ \left( \ddot\zeta - \dot\zeta \left( \dot \nu_1 +\dot\nu \right) \right) \e^{ -2\nu}
+ \left\{ \zeta'' - \zeta' \left(\nu_1' +\nu' \right) \right\} \e^{-2\, \nu_1} \right\}\pm2 \e^{-\nu_1-\nu} \left( \dot\zeta \nu' + \zeta' \dot\nu_1 -\dot\zeta' \right)\,.
\]
Should the value be positive (negative), it indicates that the GW travels faster/slower than the speed of light. As denoted in Eq.~(\ref{dis2}), a positive sign is associated with the GW moving away from the black hole, whereas a sign negative indicates the GW moving towards the black hole. Thus, the velocity of the GW as it moves through the black hole is characterized by
\begin{align}
\label{ds3}
v_\mathrm{in} \equiv \e^{- \nu_1 +\nu}\left[1-4 \kappa^2 \left\{ -\ddot\zeta + \dot\zeta \left( \dot \nu_1 +\dot\nu \right) \right\} \e^{ -2\nu}
 -4 \kappa^2 \left\{ -\zeta'' + \zeta' \left(\nu_1' +\nu' \right) \right\} \e^{-2 \nu_1} + 8 \e^{-\nu_1-\nu}
\kappa^2 \left( \dot\zeta \nu' + \zeta' \dot\nu_1 -\dot\zeta' \right)\right]\, ,
\end{align}
where the velocity of GWs is identical to the speed of light, as they propagate outward from their source
\begin{align}
\label{ds44}
v_\mathrm{out} \equiv \e^{- \nu_1 +\nu}\left[1-4 \kappa^2 \left\{ -\ddot\zeta + \dot\zeta \left( \dot \nu_1 +\dot\nu \right) \right\} \e^{ -2\nu}
 -4 \kappa^2 \left\{ -\zeta'' + \zeta' \left(\nu_1' +\nu' \right) \right\} \e^{-2 \nu_1} - 8 \e^{-\nu_1-\nu}
\kappa^2 \left( \dot\zeta \nu' + \zeta' \dot\nu_1 -\dot\zeta' \right)\right] \, .
\end{align}
If $\zeta$ is a function of $r$ or $\psi$ and $\nu_1$ is a function of $t$ or $\varphi$, the following conditions are met:
 If,  $ \dot\zeta \nu' + \zeta' \dot\nu_1 -\dot\zeta'>0$ then $v_\mathrm{in}> v_\mathrm{out}$. Conversely, if $ \dot\zeta \nu' + \zeta' \dot\nu_1 -\dot\zeta'< 0$, we have  $v_\mathrm{in}< v_\mathrm{out}$.
For instance, we can examine the expression for $\zeta(t,r)$, provided in the Appendix given by Eq. (\ref{ds4}), and restate it using the variables $(\varphi,\psi)$ yields:
\begin{align}
\label{exp}
\zeta(\varphi,\psi)= 
\frac{\zeta_0 \chi^2}{\chi^2+{r_0}^2 \e^{-\frac{2\varphi}{t_0}}} \, .
\end{align}
In the scenario where $\zeta_0$ and $t_0$ are positive constants, and under the condition specified by Eq.~(\ref{exp}), we observe that $\zeta$ approaches zero as either $t$ tends toward negative infinity or $r$ approaches zero. This observation suggests that the speed at which gravitational waves propagate coincides with light either millions of years ago or inside a black hole's core. An important observation is that when $\left|t\right|>0$, we have $\frac{\dot\zeta}{\zeta_0}>0$, and if $t<\frac{t_0}{2}$, then $\frac{\ddot\zeta}{\zeta_0}<0$. Conversely, when $t>\frac{t_0}{2}$, we find $\frac{\ddot\zeta}{\zeta_0}<0$. We observe the following properties for the function $\zeta(t)$: i-$\frac{\dot{\zeta'}}{\zeta_0}>0$, is always positive \, \, ii-if $t>-\frac{t_0}{2}$, then $\frac{\dot{\zeta'}}{\zeta_0}<0$, \, \, \, iii-if $t>\frac{t_0}{2}$, then $\frac{\dot{\zeta'}}{\zeta_0}>0$, \, \, \, iv-if $t<\frac{t_0}{2}$, then $\frac{\dot{\zeta'}}{\zeta_0}>0$. Conversely, we discover for the first black hole we get:
\begin{align}
\dot\nu_1 =&\, \frac{\dot M(t)}{r(1-\frac{2M}{r})}>0\,, \nonumber \\
\nu_1' =&\, -\frac{M(t)}{r^2(1-\frac{2M}{r})}<0\,,
\end{align} and for the second black hole we get:
\begin{align}
\dot\nu_1 =&\, \frac{1}{2t}>0\,, \nonumber \\
\nu_1' =&\, -\frac{r_0}{2r^2(1-\frac{r_0}{r})}<0\,,
\end{align}
and finally, for the third black hole, we get:
\begin{align}
\dot\nu_1 =&\, \frac{-r_0}{2rt_0(1+\frac{tr_0}{rt_0})}>0\,, \nonumber \\
\nu_1' =&\, \frac{r_0(t+t_0)}{2(r-r_0)(rt_0+tr_0)}>0\,,
\end{align}
provided that $r_0<0$.

Discussing the general choice of parameters and regions is a complex task and not very beneficial, so we can focus on some specific simplified cases instead.
\begin{itemize}
\item Initially, we examine the region near the horizon where $r_0\approx 2M(t)$. Subsequently, in Eqs.~(\ref{ds3}) and (\ref{ds44}), the terms involving $\nu_1$ and $\nu$ may dominate. However, due to the conditions $\e^{2\nu}=\e^{-2\nu_1} =0.$
Hence, the prevailing terms may be expressed as:
\begin{align}
\label{ds5}
v_\mathrm{in} \sim v_\mathrm{out} \sim c \left[ 1 - 4\,\kappa^2 \dot\zeta \left( \dot \nu_1 +\dot\nu \right) \right]\, .
\end{align}
Importantly, the distinction in velocities between inward-propagating and outward-propagating gravitational waves diminishes. If $\zeta_0 < 0$, we find that  $\dot\zeta (\dot \nu_1 + \dot\nu) < 0$, leading to the GW's propagation speed exceeding that of light.

\item Additionally, we examine the region where $r_0$ significantly exceeds $2M(t)$. Subsequently, we make the following observations:
\begin{align}
\label{ds6}
v_\mathrm{in} \sim v_\mathrm{out} \sim c \left[ 1 + 4\kappa^2 \zeta'' \right]\, .
\end{align}
Once again,  the difference in velocities between GWs moving inward and those moving outward. When $\zeta_0>0$,  we observe $\zeta''<0$,  in the region, resulting in a decrease in the GW's propagation speed, which falls below the speed of light once again. Conversely, if $\zeta_0<0$, the propagation speed exceeds the speed of light.
\end{itemize}
Normally, the speed of gravitational waves coming into a black hole differs from the speed of the wave coming out. However, these inconsistencies arise at a near second-ranking level.
 GW's can be neglected. However, a distinction may emerge in the intermediate region. It is crucial to mention that in both regions, when $\zeta_0<0$, the propagation speed of gravitational waves exceeds the speed of light. On the other hand, if $\zeta_0<0$, the propagation speed falls below the speed of light.
{ The discussion above indicates that the propagation speed of gravitational waves in Einstein-Gauss-Bonnet gravity, coupled with two scalar fields, generally differs from the speed of light. This discrepancy suggests that black holes predicted by this theory may be incompatible with the observations from GW170817. In other words, unless a novel scenario is proposed to reconcile these differences, black holes in this theory may not represent realistic astrophysical objects.

Now, let's analyze the implications of a gravitational wave propagation speed that differs from the speed of light in the context of black hole construction. For a fixed frequency, if the propagation speed is higher (lower) than the speed of light, the wavelength becomes longer (shorter). While a longer wavelength could be associated with lower frequencies, which are harder to detect, it may also imply changes in the distribution and detectability of primordial gravitational waves depending on the energy spectrum. Thus, a higher (lower) propagation speed might influence the characteristics of the primordial gravitational wave background, though not necessarily its abundance.

Another consideration is the interaction between different modes of gravitational waves and the black hole horizon. In modified gravity theories, the propagation speed of gravitational waves, if different from that of light or scalar fields, can influence the dynamics of gravity-related fluctuations. This could affect the propagation of tensor modes and potentially leave imprints on cosmological observations, such as B-mode polarization in the Cosmic Microwave Background.}
\section{Conclusion}
Black holes, once considered bizarre solutions to Einstein's equations and nowadays play a central role in astrophysics, cosmology, and foundational physics. Recent radio images of supermassive black holes provide valuable insights into matter dynamics near event horizons. While deviations from GR are possible, identifying the correct alternative theory remains challenging. In this study, we explore the dynamical black holes on the frame of GR Gauss-Bonnet theory with two scalar fields.

{ In this study, we provided three ansatzes to create time-dependent black holes with visible horizons that increase in time based on the Misner-Sharp-Hernandez mass. This horizon will appear identical to all observers associated with a spherically symmetric foliation \cite{Faraoni:2016xgy}, but it will differ for observers moving relative to the original frame in a way that breaks this symmetry, such as by a Lorentz boost in a non-radial direction \cite{Wald:1991zz,Schnetter:2005ea}.}

Our models, when constructed, may incorporate ghosts. The presence of ghosts can be mitigated by imposing constraints, akin to the mimetic constraint represented by (\ref{lambda2}). These constraints have been applied to various spacetime scenarios, as demonstrated in the referenced paper by Nojiri et al. (for example, see \cite{Nojiri:2023dvf}). Evidence indicates that the conventional cosmological solutions and self-gravitating entities such as planets, the Sun, and various types of stars within Einstein's gravitational framework also qualify as solutions in this particular model. We explored the propagation of high-frequency gravitational waves by selecting the GB coupling described in equation (\ref{ds4}).
According to the research in \cite{Oikonomou:2020sij, Nojiri:2023jtf}, the propagation speed changes due to the coupling effect during the black hole formation process. Typically, the velocity of gravitational waves entering a black hole is not the same as the velocity of the wave exiting, but these discrepancies occur at a next-to-leading order level.  Through our investigation of speed expressions, we have identified conditions under which the propagation speed remains below the speed of light and adheres to causality.

{ It is essential to emphasize that the current formulation highlights the non-dynamical nature of the scalar fields.
The scalar fields act like fluids with energy density and pressure, although they are not actual material fluids in the conventional sense. The scalar fields do not vary, so unlike regular fluids, there is no fluctuation in sound speed. In the standard fluid, breaking energy conditions can result in a sound speed exceeding the speed of light, leading to potential causality violations. However, our study does not experience this breakdown.  The lack of variation shows that the solution remains stable despite not following the energy conditions. The restrictions utilized in this study mirror the constraint used in the frame of mimetic theory \cite{Chamseddine:2013kea}, where there seems to be a type of effective dark matter that acts like dust matter but lacks pressure.   The active dark matter is not tangible but non-physical and shows no variations in spatial distribution. In addition, functional dark matter does not undergo collapse because of gravity.   Intriguingly, dynamic black holes like those analyzed in this study could offer a different option to inflation for the early Universe era.  The specifics of this early Universe black holes period will be talked about in another place.}

{ So far, there have not been many physically plausible solutions to the field equations of different theories that directly describe dynamic black holes; instead, naked singularities and wormholes are more frequently observed \cite{Vbook,Fisher:1948yn,Bergmann:1957zza,Janis:1968zz, Buchdahl:1972sj,Wyman:1981bd,Bronnikov:1973fh,Campanelli:1993sm, Vanzo:2012zu,Faraoni:2018mes,Fonarev:1994xq,Kastor:2016cqs, Faraoni:2017afs,Banijamali:2019gry,Faraoni:2017ecj}. Another issue is that certain analytical solutions become complicated when described in terms of the surface area radius, and at that time it is difficult to find the apparent horizons analytically or their expressions are not straightforward  \cite{Vbook}. In general, it is challenging to design black holes that depend on time, and we have deciphered the connection functions of gravity theories that have specified apparent horizons as their solutions.}

It will be useful to do such a procedure for regular dynamical black holes to see what is the effect of the GW. This will be done elsewhere in our future study.

\appendix
\section{The explicate form of the scalar fields}\label{Sec:App_1}
\subsection{The scalar fields of the dynamical first black hole}
\begin{align}
&\beta(t,r)=\left[ -3{r_0}^{4}\left\{  \left( r-2M \left( t \right)  \right) {t_0}^{2}{r}^{4} \left\{ {r}^{3}+\frac{16}3 {\kappa}^{2}\xi_0r-{\frac {32}{3}}{\kappa}^{2}\xi_0 M \left( t \right) \right\} \ddot M \left( t \right) +4{t_0}^{2 }{r}^{4} \left( {r}^{3}+\frac{8}3{\kappa}^{2}\xi_0 r-\frac{16}3{\kappa}^{ 2}\xi_0 M \left( t \right)  \right)  \dot M^2 \left( t \right)\right.\right.\nonumber\\
  &\left.\left.+{\frac {64}{3}}\xi_0 \left( r- 2M \left( t \right)  \right) t_0{r}^{4} {\kappa}^{2} \left( r-\frac{3}2M \left( t \right)  \right) \dot M \left( t \right) +{\frac {32}{3}}\xi_0 \left( r-2M \left( t \right)  \right) ^{2}{\kappa}^{2} \left( {r}^{4}+\frac{1}2{t_0}^{2}{r}^{2}-4M \left( t \right) {t_0}^{2}r+6  M^2 \left( t \right)  {t_0}^{2} \right) M \left( t \right)  \right\}\right.\nonumber\\
  &\left.\times {e}^{2{\frac {t}{t_0}}} -{r}^{2} \left\{ 3 {r_0}^{2}\left[  \left( r-2M \left( t \right) \right) {t_0}^{2}{r}^{4} \left( {r}^{3}+\frac{16}3{\kappa}^{2}\xi_0r-{\frac {32}{3}}{\kappa}^{2}\xi_0M \left( t \right) \right) \ddot M \left( t \right) +4{t_0}^{2 }{r}^{4} \left( {r}^{3}+\frac{8}3{\kappa}^{2}\xi_0r-\frac{16}3{\kappa}^{ 2}\xi_0M \left( t \right)  \right)  \dot M^2 \left( t \right)\right.\right.\right.\nonumber\\
  &\left.\left.\left. -{\frac {64}{3}} \left( r-\frac{5}2M \left( t \right)  \right) \xi_0 \left( r-2M \left( t \right) \right) t_0 {r}^{4}{\kappa}^{2}\dot M \left( t \right) -{\frac {32}{3}}\xi_0 \left( r-2M \left( t \right)  \right) ^{2}{\kappa}^{2}M \left( t \right)  \left( {r}^{4}-\frac{13}2{t_0}^{2 }{r}^{2}+28M \left( t \right) {t_0}^{2}r \right.\right.\right.\right.\nonumber\\
  &\left.\left.\left.\left.-30 M^2 \left( t \right){t_0}^{2} \right)  \right]{e }^{4{\frac {t}{t_0}}}+ \left(  \left( r-2M \left( t \right) \right) \ddot M \left( t \right) +4 \dot M^2 \left( t \right)  \right) {t_0}^{2}{ r}^{3} \left( {r_0}^{6}+{r}^{6}{e}^{6{\frac {t}{t_0}}} \right)  \right\}  \right] \left[{r}^{5}{\kappa}^{2} \left( r-2M \left( t \right)  \right) ^{2}{t_0}^{2} \left( {r}^{2}{e}^{2{\frac { t}{t_0}}}+{r_0}^{2} \right) ^{3}\right]^{-1}\,,\nonumber\\
 & \gamma(t,r)=\left[ 6{r_0}^{4} \left\{  \left[ {r}^{3}+\frac{16}3{\kappa}^{2}\xi_0r-16{\kappa}^{2}\xi_0M \left( t \right)  \right] t_0\dot M \left( t \right) +{\frac {32}{3}}M \left( t \right) \xi_0 \left( r-\frac{5}2M \left( t \right)  \right) { \kappa}^{2} \right\} {e}^{2{\frac {t}{t_0}}}+2r \left\{ 3r{r_0}^{2} \left[  \left( {r}^{3}+\frac{16}3{\kappa}^{2}\xi_0r\right.\right.\right.\right.\nonumber\\
 &\left.\left.\left.\left.-16 {\kappa}^{2}\xi_0M \left( t \right)  \right) t_0\dot M \left( t \right) -{\frac {32}{3}}M \left( t \right) \left( r-\frac{3}2M \left( t \right)  \right) \xi_0{\kappa}^{2} \right] {e}^{4{\frac {t}{t_0}}}+t_0  \dot M \left( t \right)    \left( {r_0}^{6}+{r}^{6}{e}^{6{ \frac {t}{t_0}}} \right)  \right\}  \right]\left[ {r}^{2}{t_0}{\kappa}^{2} \left( r-2M \left( t \right)  \right)\left( {r }^{2}{e}^{2{\frac {t}{t_0}}}\right.\right.\nonumber\\
 &\left.\left.+{r_0}^{2} \right) ^{3}\right]^{-1}\,,\nonumber\\
 &\alpha(t,r)=\left[ 3{r_0}^{4} \left\{ {r}^{4} \left( {r}^{3}+\frac{16}3{\kappa}^{2}\xi_0r-{\frac {32}{3}}{\kappa}^{2}\xi_0M \left( t \right)  \right) {t_0}^{2} \left( r-2M \left( t \right) \right) \dot M \left( t \right) +4{r}^{4} \left( {r}^{3}-\frac{16}3{\kappa}^{2}\xi_0M \left( t \right) +\frac{8}3 {\kappa}^{2}\xi_0r \right) {t_0}^{2} \dot M^2 \left( t \right)\right.\right.\nonumber\\
 &\left.\left. +{\frac {64}{3}}{r}^{4}t_0 \left( r-\frac{3}2M \left( t \right)  \right) \xi_0 \left( r-2M \left( t \right)  \right) {\kappa}^{2}\dot M \left( t \right) +32 \left( {r}^{4}+1/2{t_0}^{2}{r}^{2}-\frac{8}3M \left( t \right) {t_0}^{2}r+\frac{10}3 \left( M \left( t \right) \right) ^{2}{t_0}^{2} \right) M \left( t \right) \xi_0{\kappa}^{2} \right.\right.\nonumber\\
 &\left.\left.\left( r-2M \left( t \right)  \right) ^{2} \right\} {e}^ {2{\frac {t}{t_0}}}+{r}^{2} \left\{ 3{r_0}^{2} \left[ {r} ^{4} \left( {r}^{3}+\frac{16}3{\kappa}^{2}\xi_0r-{\frac {32}{3}}{ \kappa}^{2}x_0M \left( t \right)  \right) {t_0}^{2} \left( r-2M \left( t \right)  \right) \ddot M \left( t \right) +4{r}^{4}{t_0}^{2} \left( {r}^{3}-\frac{16}3{\kappa}^{2}x_0M \left( t \right)+\right.\right.\right.\right.\nonumber\\
 &\left.\left.\left.\left. \frac{8}3{\kappa}^{2}\xi_0r \right) \dot M^2 \left( t \right)-{ \frac {64}{3}}{r}^{4}t_0\xi_0 \left( r-\frac{5}2M \left( t \right)  \right)  \left( r-2M \left( t \right)  \right) {\kappa}^{2 } \dot M \left( t \right) -32 \left( {r}^{4}-\frac{7}6{t_0 }^{2}{r}^{2}+\frac{16}3M \left( t \right) {t_0}^{2}r-6 M^2 \left( t \right) {t_0}^{2} \right) M \left( t \right) \xi_0 \right.\right.\right.\nonumber\\
 &\left.\left.\left.\left( r-2M \left( t \right)  \right) ^{2}{ \kappa}^{2} \right] {e}^{4{\frac {t}{t_0}}}+{r}^{3} \left( {r_0}^{6}+{r}^{6}{e}^{6{\frac {t}{t_0}}} \right) {t_0}^ {2} \left\{  \left( r-2M \left( t \right)  \right) \ddot M \left( t \right) +4  \dot M^2 \left( t \right) \right\}  \right\}  \right] \left[{r}^{3}{t_0}^{2} \left( r-2M \left( t \right)  \right) ^{4}{\kappa}^{2} \left( { r}^{2}{e}^{2{\frac {t}{t_0}}}+{r_0}^{2} \right) ^{3}\right]^{-1} \,,\nonumber\\
 &\Phi(t,r)=2{r_0}^{2}\xi_0 M \left( t \right) \left[ - \left\{ {r}^{4} \dot M \left( t \right) t_0+ \left( {r}^{4}+\frac{1}2{t_0}^{2}{r}^{2}-3M \left( t \right) {t_0}^{2}r+4 \left( M \left( t \right)  \right) ^{2}{t_0}^{2} \right)  \left( r-2M \left( t \right)  \right)  \right\} {r_0}^{2}{e}^{2{\frac {t}{t_0}}}+{r}^{2}{e}^{4{\frac {t}{t_0}}} \left\{ \right.\right.\nonumber\\
 &\left.\left. \times\left( {r }^{4}-\frac{5}2{t_0}^{2}{r}^{2}-{r}^{4} \dot M \left( t \right)  t_0+11M \left( t \right) {t_0}^{2} r-12  M ^2\left( t \right) {t_0}^{2} \right) \left( r-2M \left( t \right)  \right)  \right\}  \right] \left[{r}^{4}{t_0}^{2} \left( r-2M \left( t \right) \right) ^{2} \left( {r}^{2}{e}^{2{\frac {t}{t_0}}}+{r_0} ^{2} \right) ^{3}\right]^{-1}\,,
\end{align}
where we have assumed the form the Gauss-Bonnet coupling has the form
\begin{align}
\label{ds4}
\xi(t,r)= \frac{\xi_0 r^2}{r^2+{r_0}^2 \e^{-\frac{2t}{t_0}}} \, .
\end{align}

\subsection{The scalar fields of the dynamical second black hole}
\begin{align}
&\beta(t,r)=\left[ 3 \left\{  \left[ -{t_0}^{3}t+ \left( {t}^{2}+\frac{8}3{\kappa}^{2}\xi_0 \right) {t_0}^{2}-8{\kappa}^{2}t\xi_0 t_0-\frac{8}3{\kappa}^{2}{t}^{2}\xi_0 \right] {r}^{5}- \left( t {t_0}^{2}+\frac{16}3{\kappa}^{2}t\xi_0-\frac{8}3t_0\xi_0 {\kappa}^{2} \right) r_0 \left( t-t_0\right) {r}^{4}\right.\right.\nonumber\\
&\left.\left.+\frac{16} 3{t_0}^{3}{\kappa}^{2}\xi_0 \left( t-t_0 \right) {r }^{3}-8 \left( t-\frac{2}3t_0\right) \xi_0{t_0}^{3}{ \kappa}^{2}r_0{r}^{2}+\frac{8}3{r_0}^{2}{t_0}^{3}{\kappa} ^{2}\xi_0 \left( 3t_0+t \right) r-8{t_0}^{4}{r_0}^{3}{\kappa}^{2}\xi_0 \right\} {r_0}^{4}{e}^{2{\frac {t} {t_0}}}\right.\nonumber\\
&\left.+ \left\{ 3 \left[  \left\{  \left( {t}^{ 2}+\frac{8}3{\kappa}^{2}\xi_0 \right) {t_0}^{2}-{t_0}^{3}t+{\frac {40}{3}} {\kappa}^{2}t\xi_0t_0-\frac{8}3{\kappa}^{2}{t}^{2}\xi_0 \right\} {r}^{5}- \left[ \left( {t}^{2}+\frac{8}3{\kappa }^{2}\xi_0 \right) {t_0}^{2} -{t_0}^{3}t+{\frac {40}{3}}{\kappa}^{2}t\xi_0t_0-\frac{16}3{\kappa}^{2}{t}^{2}\xi_0 \right] r_0 {r}^{4}\right.\right.\right.\nonumber\\
&\left. \left.\left.-16{t_0}^{3}{\kappa}^{2}\xi_0 \left( -t_0+t \right) {r}^{3}+{\frac {104}{3}}r_0{t_0}^{3}{\kappa}^{2 }\xi_0 \left( -2t_0+t \right) {r}^{2}-{\frac {56}{3}}{r_0}^{2}{t_0}^{3}{\kappa}^{2}\xi_0 \left( t-5t_0 \right) r-40{t_0}^{4}{r_0}^{3}{\kappa}^{2}\xi_0 \right] {r_0}^{2}{e}^{4{\frac {t}{t_0}}}\right.\right.\nonumber\\
& \left.\left.+t{t_0}^{2} \left( r-r_0 \right)  \left( {r_0}^{6}+{r}^{6}{e}^{6{ \frac {t}{t_0}}} \right)  \left( -t_0+t \right)  \right\} {r} ^{2} \right]\left[ {r}^{5}{t_0}{\kappa}^{2}{t}^{3} \left( {r}^{ 2}{e}^{2{\frac {t}{t_0}}}+{r_0}^{2} \right) ^{3}\right]^{-1} \,,\nonumber\\
&\gamma=\left[ 3 \left\{ -{\frac {40}{3}}\xi_0t_0 \left(t -\frac{3}5t_0\right) {\kappa}^{2}{r_0}^{2}- \left\{ {r}^{2}tt_0+{\frac {40}{3}}\xi_0\left( {t}^{2}+\frac{6}5{t_0}^{2}-2 tt_0 \right) {\kappa}^{2} \right\} rr_0+ \left[ {r}^{2}tt_0+{\frac {32}{3}}\xi_0\left( {t}^{2}+\frac{3}4{t_0}^{2 }-\frac{5}4tt_0 \right) {\kappa}^{2} \right] {r}^{2} \right\}\right.\nonumber\\
& \left. \times{r_0}^{4}{e}^{2{\frac {t}{t_0}}}+ \left\{ 3 \left[ 8t_0\xi_0{\kappa}^{2} \left( t_0+t \right) {r_0}^{2}- \left( {r}^{2}tt_0-8{\kappa}^{2}\xi_0\left( -2tt_0+{t}^{2}-2{t_0}^{2} \right)  \right) rr_0+{r}^{2} \left( {r}^{2}tt_0-{\frac {32}{3}}\xi_0\left( {t}^{2}-\frac{3}4tt_0-\frac{3}4{t_0}^{2} \right) {\kappa}^{2} \right) \right]\right.\right.\nonumber\\
&\left. \left. r{r_0}^{2}{e}^{4{\frac {t}{t_0}}}+tt_0 \left( r-r_0 \right)  \left( {r_0}^{6}+{r}^{6}{e}^{6{ \frac {t}{t_0}}} \right)  \right\} r \right]\left[ {r}^{2}{t}^{2} \left( r-r_0 \right) {\kappa}^{2}{t_0} \left( {r }^{2}{e}^{2{\frac {t}{t_0}}}+{r_0}^{2} \right) ^{3}\right]^{-1}\,,\nonumber\\
&\alpha(t,r)=\left[ 3{r_0}^{4} \left\{ {r}^{4} \left( {r}^{3}+\frac{16}3{\kappa}^{2}\xi_0r-{\frac {32}{3}}{\kappa}^{2}\xi_0M \left( t \right)  \right) {t_0}^{2} \left( r-2M \left( t \right) \right) \ddot M \left( t \right) +4{r}^{4} \left( {r}^{3}-\frac{16}3{\kappa}^{2}\xi_0M \left( t \right) +\frac{8}3 {\kappa}^{2}\xi_0r \right) {t_0}^{2} \dot M^2 \left( t \right)\right.\right.\nonumber\\
&\left.\left.+{\frac {64}{3}}{r}^{4}t_0 \left[ r-\frac{3}2M \left( t \right)  \right] \xi_0 \left( r-2M \left( t \right)  \right) {\kappa}^{2}\dot M \left( t \right) +32 \left[ {r}^{4}+\frac{{t_0}^{2}{r}^{2}}2-\frac{8}3M \left( t \right) {t_0}^{2}r+\frac{10}3 \left( M \left( t \right) \right) ^{2}{t_0}^{2} \right]M \left( t \right) \xi_0 \left( r-2M \left( t \right)  \right) ^{2}{\kappa}^{2} \right\}\right.\nonumber\\
&\left. \times{e}^ {2{\frac {t}{t_0}}}+{r}^{2} \left( 3{r_0}^{2} \left( {r} ^{4} \left[ {r}^{3}+\frac{16}3{\kappa}^{2}\xi_0r-{\frac {32}{3}}{ \kappa}^{2}\xi_0M \left( t \right)  \right] {t_0}^{2} \left( r-2M \left( t \right)  \right) \ddot M \left( t \right) +4{r}^{4} \left( {r}^{3}-\frac{16}3{\kappa}^{2}\xi_0M \left( t \right) +\frac{8}3{\kappa}^{2}\xi_0r \right) {t_0}^{2} \dot M^2\left( t \right)\right.\right.\right.\nonumber\\
&\left.\left.\left. -{ \frac {64}{3}}{r}^{4}t_0\xi_0 \left( r-\frac{5}2M \left( t \right)  \right)  \left( r-2M \left( t \right)  \right) {\kappa}^{2 }\dot M \left( t \right) -32 \left( {r}^{4}-\frac{7}6{t_0 }^{2}{r}^{2}+\frac{16}3M \left( t \right) {t_0}^{2}r-6 \left( M \left( t \right)  \right) ^{2}{t_0}^{2} \right) M \left( t \right) \xi_0 {e}^{4{\frac {t}{t_0}}}\right.\right.\right.\nonumber\\
&\left.\left.\left. \times\left( r-2M \left( t \right)  \right) ^{2}{ \kappa}^{2} \right)+{r}^{3} \left( {r_0}^{6}+{r}^{6}{e}^{6{\frac {t}{t_0}}} \right) {t_0}^ {2} \left(  \left( r-2M \left( t \right)  \right) \ddot M \left( t \right) +4\dot M^2 \left( t \right) \right)  \right)  \right]\left[ {r}^{3}{t_0}^{ 2} \left( r-2M \left( t \right)  \right) ^{4}{\kappa}^{2} \left( { r}^{2}{e}^{2{\frac {t}{t_0}}}+{r_0}^{2} \right) ^{3}\right]^{-1}\,,
\nonumber\\
&\Phi=32{r_0}^{2}\xi_0\left[ - \left\{ {r}^{4} \dot M \left( t \right) t_0+ \left( {r}^{4}+\frac{1}2{t_0}^{2}{r}^{2}-3M \left( t \right) {t_0}^{2}r+4 M^2 \left( t \right) {t_0}^{2} \right)  \left( r-2M \left( t \right)  \right)  \right\} {r_0}^{2}{e}^{2{\frac {t}{t_0}}}+{r}^{2}{e}^{4{\frac {t}{t_0}}} \left( -{r}^{4} \dot M \left( t \right) t_0\right.\right.\nonumber\\
&\left.\left.+ \left( {r }^{4}-\frac{5}2{t_0}^{2}{r}^{2}+11M \left( t \right) {t_0}^{2} r-12 \left( M \left( t \right)  \right) ^{2}{t_0}^{2} \right) \left( r-2M \left( t \right)  \right)  \right)  \right] M \left( t \right) \left[{r}^{4}{t_0}^{2} \left( r-2M \left( t \right) \right) ^{2} \left( {r}^{2}{e}^{2{\frac {t}{t_0}}}+{r_0} ^{2} \right) ^{3}\right]^{-1}\,,
\end{align}
where we have used the coupling Gauss-Bonnet given by Eq.~(\ref{ds4}).

\subsection{The scalar fields of the dynamical third black hole}
\begin{align}
&\beta(t,r)=
\left[ 3{r_0}^{4} \left\{ \frac{1}8 \left( r_0 \left[ 8 \left( \frac{16}3{\kappa}^{2}\xi_0-{r}^{2} \right) t^2r_0^3+ \left(16{t}^{2}{ r}^{3}-96r{\kappa}^{2}\xi_0{t}^{2}-r^5 \right) r_0^2+ \left[ r^4+ \left( \frac {32}{3} \kappa^2\xi_0-9t^2 \right) r^2+\frac {160}{3}\kappa^2t^2\xi_0 \right] r^2r_0\right.\right.\right.\right.\nonumber\\
&\left.\left.\left.\left.-{\frac {32}{3}}{ r}^{5}{\kappa}^{2}\xi_0 \right] {t_0}^{5/2}-8 \left[ \frac{5}2r_0 \left\{  \left( -{\frac {56}{15}}{\kappa}^{2}\xi_0+\frac{4}5 {r}^{2} \right) {r_0}^{2}+ \left( -{\frac {17}{10}}{r}^{3}+{ \frac {128}{15}}{\kappa}^{2}\xi_0r \right) r_0-{\frac {24 }{5}}{\kappa}^{2}\xi_0{r}^{2}+{r}^{4} \right\} t{t_0}^{7/2 }\right.\right.\right.\right.\nonumber\\
&\left.\left.\left.\left.+r \left\{  \left[  \left( -4{\kappa}^{2}\xi_0+5/8{r}^{2} \right) {r_0}^{2}+ \left( {\frac {28}{3}}{\kappa}^{2}\xi_0 r-\frac{3}2{r}^{3} \right) r_0+{r}^{4}-\frac{16}3{\kappa}^{2}\xi_0 {r}^{2} \right] {t_0}^{9/2}-\frac{1}8 \left( r-r_0 \right) {r_0}^{2} \left( r \left( r_0r-{\frac {64}{3}}{\kappa}^{2} \xi_0 \right) {t_0}^{3/2}\right.\right.\right.\right.\right.\right.\nonumber\\
&\left.\left.\left.\left.\left.\left.-{\frac {32}{3}}t{\kappa}^{2}\xi_0\sqrt {t_0}r_0 \right) t \right\}  \right] r \right) \sqrt {r-r_0}\sqrt {t_0r+tr_0}+ \left( r-r_0 \right)  \left[ -t \left[  \left(\frac{4}3t_0{\kappa}^{2}\xi_0 -4t{\kappa}^{2}\xi_0+t_0{t}^{2} \right) {r}^{2}+\frac{16}3 {t_0}^{2}t{\kappa}^{2}\xi_0 \right] t_0{r_0}^{4 }\right.\right.\right.\nonumber\\
&\left.\left.\left.+ \left\{  \left(  \left( -\frac{8}3{\kappa}^{2}\xi_0+{t_0}^{2} \right) {t}^{3}- \left( {\frac {20}{3}}t_0{\kappa}^{2}\xi_0+3{t_0}^{3} \right) {t}^{2}+{\frac {28}{3}}{t_0}^{2} t{\kappa}^{2}\xi_0-\frac{4}3{t_0}^{3}{\kappa}^{2}\xi_0 \right) {r}^{2}+{\frac {40}{3}} \left( t-\frac{3}5t_0\right) t{ \kappa}^{2}\xi_0{t_0}^{3} \right\} r{r_0}^{3}\right.\right.\right.\nonumber\\
&\left.\left.\left.+3{r}^{2 }t_0 \left[  \left\{  \left( -{\frac {16}{9}}{\kappa}^{2}\xi_0+{t_0}^{2} \right) {t}^{2}-\left( {\frac {40}{9}}t_0 {\kappa}^{2}\xi_0+{t_0}^{3} \right) t+{\frac {16}{9}}{t_0}^{2}{\kappa}^{2}\xi_0 \right\} {r}^{2}-\frac{8}3t{t_0}^{2} {\kappa}^{2}\xi_0 \left( -3t_0+t \right)  \right] {r_0}^{2}+3 \left\{  \left[  \left( {t_0}^{2}-{\frac {8}{9}}{ \kappa}^{2}\xi_0 \right) t\right.\right.\right.\right.\right.\nonumber\\
&\left.\left.\left.\left.\left.-\frac{1}3{t_0}^{3}-{\frac {20}{9}}t_0{\kappa}^{2}\xi_0 \right] {r}^{2}-\frac{16}3{\kappa}^{2}\xi_0 \left( t-\frac{1}3t_0 \right) {t_0}^{2} \right\} {r}^{3}{t_0}^{2}r_0+ \left( -\frac{16}3{\kappa}^{2}\xi_0+{r}^{2} \right) {r}^{4}{t_0}^{5} \right]  \right\} {e}^{2{\frac {t}{t_0}}}+3{r_0}^{2}{r}^{2} \left[ \frac{1}8\sqrt {r-r_0}\right.\right.\nonumber\\
&\left.\left. \left[  \left\{ \left( {\frac {736}{3}}r{\kappa}^{2}\xi_0{t}^{2}-{r}^{5}+16{t}^{2}{r}^{3} \right) {r_0}^{2} -8{t}^{2} \left( 16{\kappa}^{2}\xi_0+{r}^{2} \right) {r_0}^{3}+{r}^{2 } \left\{ {r}^{4}+ \left( {\frac {32}{3}}{\kappa}^{2}\xi_0-9{t} ^{2} \right) {r}^{2}-{\frac {352}{3}}{\kappa}^{2}{t}^{2}\xi_0 \right\} r_0-{\frac {32}{3}}{r}^{5}{\kappa}^{2}\xi_0 \right\} \right.\right.\right.\nonumber\\
&\left.\left.\left.\times r_0{t_0}^{5/2}-8 \left\{ \frac{5}2 \left[  \left( { \frac {40}{3}}{\kappa}^{2}\xi_0+\frac{4}5{r}^{2} \right) {r_0}^ {2}- \left( {\frac {17}{10}}{r}^{3}+{\frac {128}{5}}{\kappa}^{2}\xi_0r \right) r_0+{\frac {184}{15}}{\kappa}^{2}\xi_0 {r}^{2}+{r}^{4} \right] r_0t{t_0}^{7/2}+ \left[  \left( \left( {\frac {52}{3}}{\kappa}^{2}\xi_0+\frac{5}8{r}^{2} \right) {r_0}^{2}\right.\right.\right.\right.\right.\right.\nonumber\\
&\left.\left.\left.\left.\left.\left.- \left\{ {\frac {100}{3}}{\kappa}^{2}\xi_0r+\frac{3}2 {r}^{3} \right\} r_0+{r}^{4}+16{\kappa}^{2}\xi_0{r}^{2} \right) {t_0}^{9/2}-\frac{1}8 \left( r-r_0 \right) {r_0}^{ 2} \left\{ r \left( r_0r-{\frac {64}{3}}{\kappa}^{2}\xi_0 \right) {t_0}^{3/2}-{\frac {32}{3}}t{\kappa}^{2}\xi_0 \sqrt {t_0}r_0 \right\} t \right] r \right\} r \right]\right.\right.\nonumber\\
&\left.\left. \sqrt { t_0r+tr_0}+ \left[ - \left\{  \left( {\frac {20}{3}}t{ \kappa}^{2}\xi_0+\frac{4}3t_0{\kappa}^{2}\xi_0+t_0{t }^{2} \right) {r}^{2}-16{t_0}^{2}t{\kappa}^{2}\xi_0 \right\} tt_0{r_0}^{4}+r \left[  \left\{  \left( \frac{8}3{ \kappa}^{2}\xi_0+{t_0}^{2} \right) {t}^{3}+ \left( {\frac {28 }{3}}t_0{\kappa}^{2}\xi_0-3{t_0}^{3} \right) {t}^{ 2}\right.\right.\right.\right.\right.\nonumber\\
&\left.\left.\left.\left.\left.-12{t_0}^{2}t{\kappa}^{2}\xi_0-\frac{4}3{t_0}^{3}{\kappa }^{2}\xi_0 \right\} {r}^{2}-24t{\kappa}^{2}\xi_0{t_0}^ {3} \left( t-\frac{5}3t_0 \right)  \right] {r_0}^{3}+3 \left[  \left\{  \left( {\frac {16}{9}}{\kappa}^{2}\xi_0+{t_0}^{2} \right) {t}^{2}+ \left( {\frac {56}{9}}t_0{\kappa}^{ 2}\xi_0-{t_0}^{3} \right) t-{\frac {16}{9}}{t_0}^{2}{ \kappa}^{2}\xi_0 \right\} {r}^{2}\right.\right.\right.\right.\nonumber\\
&\left.\left.\left.\left.\left.-{\frac {16}{9}}{t_0}^{2}{ \kappa}^{2}\xi_0 \right\} {r}^{2}+8/3{\kappa}^{2} \left\{ -{\frac {23}{3}}tt_0+10/3{t_0}^{2}+{t}^{2} \right\} \xi_0{{ \it t0}}^{2} \right] {r}^{2}t_0{r_0}^{2}+3 \left[ \left\{  \left( {t_0}^{2}+{\frac {8}{9}}{\kappa}^{2}\xi_0 \right) t-1/3{t_0}^{3}+{\frac {28}{9}}t_0{\kappa}^{2} \xi_0 \right\} {r}^{2}\right.\right.\right.\right.\nonumber\\
&\left.\left.\left.\left. +{\frac {64}{9}}{t_0}^{2}{\kappa}^{2}\xi_0\left( t-2t_0 \right)  \right] {r}^{3}{t_0}^{2 }r_0+{r}^{4}{t_0}^{5} \left( 16{\kappa}^{2}\xi_0+{r}^{ 2} \right)  \right]  \left( r-r_0 \right)  \right] {e}^{4{ \frac {t}{t_0}}}+ \left( \frac{1}8 \left[ {r_0}^{2} \left\{{r}^{4}  -8{ t}^{2}{r_0}^{2}- \left( {r}^{3}-16r{t}^{2} \right) r_0-9 {t}^{2}{r}^{2}\right\} {t_0}^{5/2}\right.\right.\right.\nonumber\\
&\left.\left.\left.-8r \left\{ \frac{5}2 \left( \frac{4}5{r_0}^{2}+{r}^{2}-{\frac {17}{10}}r_0r \right) r_0t{t_0}^{7/2}+r \left(  \left( -\frac{3}2r_0 r+{r}^{2}+\frac{5}8{r_0}^{2} \right) {t_0}^{9/2}-\frac{1}8{r_0} ^{3}{t_0}^{3/2}t \left( r-r_0 \right)  \right)  \right\} \right] \sqrt {r-r_0}\sqrt {t_0r+tr_0}\right.\right.\nonumber\\
&\left.\left.+{t_0}^{2 } \left( r-r_0 \right) ^{2} \left( t_0r+tr_0 \right) ^{3} \right)  \left( {r_0}^{6}+{r}^{6}{e}^{6{\frac {t}{t_0 }}} \right)  \right] \left[ {t_0}^{3/2}\left( r-r_0 \right) ^{3/ 2} \left( t_0r+tr_0 \right) ^{7/2}{r}^{-2}{\kappa}^{2} \left( {r}^{2}{e}^{2{\frac {t}{t_0}}}+{r_0}^{2} \right) ^ {3}\right]^{-1}\,,\nonumber\\
&\gamma(t,r)=\frac{r_0}2 \left[ 3{r_0}^{3} \left\{ -{\frac {64}{3}}\sqrt {r-r_0} \left(  \left( {r}^{2}-3/2r_0r+3/8{r_0}^{2} \right) {t_0}^{3/2}+{\frac {7}{8}} \left( r-{\frac {8}{7}}{ \it r0} \right) \sqrt {t_0}r_0t \right) \xi_0{\kappa }^{2}\sqrt {t_0r+tr_0}+ \left( t_0r+tr_0 \right)\right.\right.\nonumber\\
&\left.\left.  \left[  \left\{ {\frac {64}{3}}\xi_0  \left( \frac{t_0}8-t \right) {\kappa}^{2}-t_0{r}^{2} \right\} {r_0}^{2} + \left( t_0{r}^{2}+{\frac {56}{3}} \left( t-{\frac {10}{7}} t_0\right) \xi_0{\kappa}^{2} \right) rr_0+{\frac { 64}{3}}{\kappa}^{2}\xi_0t_0{r}^{2} \right]  \right\} {e} ^{2{\frac {t}{t_0}}}+3{r}^{2}r_0 \left( {\frac {64}{3} }\sqrt {r-r_0}\xi_0{\kappa}^{2}\right.\right.\nonumber\\
&\left.\left.  \left(  \left( {r}^{2}-\frac{r_0}2r-3/8{r_0}^{2} \right) {t_0}^{3/2}+{\frac {9} {8}} \left( r-{\frac {8}{9}}r_0 \right) \sqrt {t_0}r_0t \right) \sqrt {t_0r+tr_0}+ \left( t_0r+tr_0 \right)  \left[  \left( {\frac {64}{3}}\xi_0 \left( \frac{1}8t_0+t \right) {\kappa}^{2}-t_0{r}^{2} \right) {r_0}^{2}\right.\right.\right.\nonumber\\
&\left.\left.\left.+r \left[ t_0{r}^{2}-24\xi_0 \left(t -\frac{2}3t_0 \right) {\kappa}^{2} \right] r_0-{\frac {64}{3}}{\kappa}^ {2}\xi_0t_0{r}^{2} \right]  \right){e}^{4{\frac {t}{t_0}}}+t_0 \left( r-r_0 \right)  \left( {r_0}^{6} +{r}^{6}{e}^{6{\frac {t}{t_0}}} \right)  \left( t_0r+tr_0 \right)  \right] \left[{r} \left( t_0r+tr_0 \right) ^{2}{t_0}{\kappa}^{2} \left( r-r_0 \right) \right.\nonumber\\
&\left. \left( {r}^{2}{e}^{2{\frac {t}{t_0}}}+{r_0}^{2} \right) ^{3}\right]^{-1}\,,\nonumber\\
&\alpha(t,r)=- \left[ 3r{r_0}^{4} \left\{ 1/8\sqrt {r-r_0} \left[ \left\{  \left( {r_0}^{2}-{\frac {256}{3}}{\kappa}^{2}\xi_0 \right) {r}^{5}+ \left( {\frac {224}{3}}r_0{\kappa}^{2}\xi_0-{r_0}^{3} \right) {r}^{4}-9 \left( {t}^{2}-{\frac {32}{27} }{\kappa}^{2}\xi_0 \right) {r_0}^{2}{r}^{3}+16{r}^{2}{t}^ {2}{r_0}^{3}\right)\right.\right.\right.\nonumber\\
&\left.\left.\left.\left.8{t}^{2} \left( \frac{4}3{\kappa}^{2}\xi_0-{r_0}^{2} \right) {r_0}^{2}r-{\frac {32}{3}}{r_0}^{3}{ \kappa}^{2}\xi_0{t}^{2} \right\} {t_0}^{7/2}+ \left( {\frac {32}{3}} r_0{r}^{2}{\kappa}^{2}\xi_0-{ \frac {32}{3}}{r_0}^{2}r{\kappa}^{2}\xi_0-8{r}^{5}-5{r_0}^{2}{r} ^{3}+12r_0{r}^{4} \right) {t_0}^{11/2}\right.\right.\right.\nonumber\\
&\left.\left.\left.+t \left\{  \left( -20{r}^{4}+34r_0{r}^{3}+ \left( -16{r_0}^{2}+{\frac {32}{3}}{\kappa}^{2}\xi_0 \right) {r}^{2}-{\frac {32}{3}}{r_0}^{2}{\kappa}^{2}\xi_0 \right) {t_0}^{9/2}+ \left[ \left\{  \left( {r_0}^{2}-256{\kappa}^{2}\xi_0 \right) r-{ \frac {64}{3}}r_0{\kappa}^{2}\xi_0 \right\} r{t_0}^{5 /2}\right.\right.\right.\right.\right.\nonumber\\
&\left.\left.\left.\left.\left.-256t\xi_0 \left(  \left( r+1/24r_0 \right) {t_0}^{3/2}+1/3r_0t\sqrt {t_0} \right) {\kappa}^{2}r_0 \right] r \left( r-r_0 \right)  \right\} r_0 \right] \sqrt { rt_0+tr_0}+t_0 \left[  \left( {t_0}^{5}+{\frac { 32}{3}}{t_0}^{3}{\kappa}^{2}\xi_0 \right) {r}^{5}\right.\right.\right.\nonumber\\
&\left.\left.\left.+3{t_0}^{2}r_0 \left\{  \left( {t_0}^{2}+{\frac {88}{9}}{ \kappa}^{2}\xi_0 \right) t-\frac{1}3{t_0}^{3}-{\frac {52}{9}}{ \kappa}^{2}\xi_0t_0 \right\} {r}^{4}+3t_0 \left\{ \left( {t_0}^{2}+{\frac {80}{9}}{\kappa}^{2}\xi_0 \right) {t}^{2}- \left( {t_0}^{3}+{\frac {136}{9}}{\kappa}^{2}\xi_0t_0 \right) t+{\frac {16}{9}}{t_0}^{2}{\kappa}^{2}\xi_0 \right\} {r_0}^{2}{r}^{3}\right.\right.\right.\nonumber\\
&\left.\left.\left.+ \left[  \left\{  \left( 8{\kappa} ^{2}\xi_0+{t_0}^{2} \right) {t}^{3}+ \left( -{\frac {116}{3}} {\kappa}^{2}\xi_0t_0-3{t_0}^{3} \right) {t}^{2}+{ \frac {28}{3}}{t_0}^{2}{\kappa}^{2}\xi_0t-\frac{4}3{t_0} ^{3}{\kappa}^{2}\xi_0 \right\} {r_0}^{2}-16/3{\kappa}^{2}\xi_0{t_0}^{4} \left( t+t_0 \right)  \right] r_0 {r}^{2}\right.\right.\right.\nonumber\\
&\left.\left.\left.- \left[ t \left\{  \left( {t_0}^{2}+{\frac {32}{3}}{ \kappa}^{2}\xi_0 \right) {t}^{2}-4t_0{\kappa}^{2}\xi_0t+4/3{t_0}^{2}{\kappa}^{2}\xi_0 \right\} {r_0}^{2}+8 /3{t_0}^{3}{\kappa}^{2}\xi_0 \left( t+t_0 \right) \left( t-2t_0 \right)  \right] {r_0}^{2}r+8/3t{t_0 }^{3}{r_0}^{3}{\kappa}^{2}\xi_0 \left( t+t_0 \right) \right]\right.\right.\nonumber\\
&\left.\left.  \left( r-r_0 \right)  \right\} {e}^{2{\frac {t}{t_0}}}+3{r}^{3} \left[ 1/8\sqrt {r-r_0} \left\{  \left[ \left( {r_0}^{2}+{\frac {256}{3}}{\kappa}^{2}\xi_0 \right) {r}^{5}+ \left( -96r_0{\kappa}^{2}\xi_0-{r_0}^{3} \right) {r}^{4}-9 \left( {t}^{2}-{\frac {32}{27}}{\kappa}^{ 2}\xi_0 \right) {r_0}^{2}{r}^{3}+16{r}^{2}{t}^{2}{r_0} ^{3}\right.\right.\right.\right.\nonumber\\
&\left.\left.\left.\left.-8{t}^{2} \left( -\frac{4}3{\kappa}^{2}\xi_0+{r_0}^{2} \right) {r_0}^{2}r-{\frac {32}{3}}{r_0}^{3}{\kappa}^{2}\xi_0{t}^{2} \right] {t_0}^{7/2}+ \left( -{\frac {32}{3}}{ r_0}^{2}r{\kappa}^{2}\xi_0+{\frac {32}{3}}r_0{r}^{2} {\kappa}^{2}\xi_0-8{r}^{5}-5{r_0}^{2}{r}^{3}+12r_0 {r}^{4} \right) {t_0}^{11/2}\right.\right.\right.\nonumber\\
&\left.\left.\left.+t \left[  \left\{ -20{r}^{4}+34 r_0{r}^{3}+ \left( -16{r_0}^{2}+{\frac {32}{3}}{\kappa }^{2}\xi_0 \right) {r}^{2}-{\frac {32}{3}}{r_0}^{2}{\kappa} ^{2}\xi_0 \right\} {t_0}^{9/2}+ \left( r \left(  \left( {r_0}^{2}+256{\kappa}^{2}\xi_0 \right) r-{\frac {64}{3}}r_0{\kappa}^{2}\xi_0 \right) {t_0}^{5/2}\right.\right.\right.\right.\right.\nonumber\\
&\left.\left.\left.\left.\left.+256t \left\{ \left( r-\frac{r_0}{24} \right) {t_0}^{3/2}+\frac{r_0t}{3} \sqrt {t_0} \right\} \xi_0{\kappa}^{2}r_0 \right) r \left( r-r_0 \right)  \right] r_0 \right\} \sqrt {rt_0+ tr_0}+ \left[  \left( {t_0}^{5}-{\frac {32}{3}}{t_0}^ {3}{\kappa}^{2}\xi_0 \right) {r}^{5}+3{t_0}^{2} \left\{ \left( {t_0}^{2}-{\frac {88}{9}}{\kappa}^{2}\xi_0 \right) t\right.\right.\right.\right.\nonumber\\
&\left.\left.\left.\left.-\frac{1}3 \left( -20{\kappa}^{2}\xi_0+{t_0}^{2} \right) t_0 \right\} r_0{r}^{4}+3t_0 \left[  \left( {t_0}^ {2}-{\frac {80}{9}}{\kappa}^{2}\xi_0 \right) {t}^{2}+ \left( { \frac {152}{9}}{\kappa}^{2}\xi_0t_0-{t_0}^{3} \right) t-{\frac {16}{9}}{t_0}^{2}{\kappa}^{2}\xi_0 \right] {r_0}^{2}{r}^{3}+r_0 \left[  \left\{  \left( {t_0}^{2} \right) {t}^{3}-8{ \kappa}^{2}\xi_0\right.\right.\right.\right.\right.\nonumber\\
&\left.\left.\left.\left.\left.+ \left( {\frac { 124}{3}}{\kappa}^{2}\xi_0t_0-3{t_0}^{3} \right) {t }^{2}-12{t_0}^{2}{\kappa}^{2}\xi_0t-4/3{t_0}^{3}{ \kappa}^{2}\xi_0 \right\} {r_0}^{2}-{\frac {32}{3}}{\kappa}^ {2}\xi_0{t_0}^{4} \left( t+t_0 \right)  \right] {r}^{2 }- \left[ t \left\{  \left( {t_0}^{2}-{\frac {32}{3}}{\kappa}^{2 }\xi_0 \right) {t}^{2}+{\frac {20}{3}}t_0{\kappa}^{2}\xi_0t\right.\right.\right.\right.\right.\nonumber\\
&\left.\left.\left.\left.\left.+4/3{t_0}^{2}{\kappa}^{2}\xi_0 \right\} {r_0 }^{2}+8 \left( t+t_0 \right) \xi_0{t_0}^{3} \left( t -4/3t_0 \right) {\kappa}^{2} \right] {r_0}^{2}r+8t{t_0}^{3}{r_0}^{3}{\kappa}^{2}\xi_0 \left( t+t_0 \right)  \right] t_0 \left( r-r_0 \right)  \right] {r_0}^{2}{e}^{4{\frac {t}{t_0}}}+ \left( {r}^{6}{e}^{6{\frac { t}{t_0}}}+{r_0}^{6} \right)\right.\nonumber\\
&\left.  \left\{ \frac{1}8\sqrt {r-r_0} \left[ {r_0}^{2} \left({r}^{4} -9{r}^{2}{t}^{2}-8{t}^{2}{r_0}^ {2}-r_0{r}^{3}+16r_0r{t}^{2} \right) {t_0 }^{7/2}-8r \left\{ \left(\frac{5}8{r_0}^ {2} -\frac{3}2r_0r+{r}^{2} \right) r{t_0}^{11/2}-\frac{t}8 \left[  \left(34r_0r  -16{r_0}^{ 2}-20{r}^{2} \right)\right.\right.\right.\right.\right.\nonumber\\
&\left.\left.\left.\left.\left.  {t_0}^{9/2}+ \left( r-r_0 \right) r{r_0}^{2}{t_0}^{5/2} \right] r_0 \right\}  \right] \sqrt {rt_0+tr_0}+{t_0}^{3} \left( r- r_0 \right) ^{2} \left( rt_0+tr_0 \right) ^{3} \right\} \right]\left[  \left( rt_0+tr_0 \right) ^{5/2} \left( r-r_0 \right) ^{5/2}{t_0}^{7/2}{r}^{2}{\kappa}^{2} \left( {r}^{2}{ e}^{2{\frac {t}{t_0}}}\right.\right.\nonumber\\
&\left.\left.+{r_0}^{2} \right) ^{3}\right]^{-1}\,,\nonumber\\
&\Phi(t,r)=\frac{1}2 \left[ 6{r_0}^{4} \left\{ -\frac{16}3 \left[ {\frac {9}{32}} \left( -\frac{2}3r_0+r \right) r_0t \left( {r}^{2}-\frac{8}3{ \kappa}^{2}\xi_0 \right) \left( \frac{3}{16} \left( r- \frac{1}2r_0 \right)  \left( {r}^{2}-\frac{8}3{\kappa}^{2}\xi_0 \right) {t_0}^{9/2}+ \left( 2r_0 \left( \frac{1}8r_0+r \right) t{t_0}^{3/2}\right.\right.\right.\right.\right.\nonumber\\
&\left.\left.\left.\left.\left. + \left( r+\frac{1}4r_0 \right) r{t_0} ^{5/2}+{r_0}^{2}\sqrt {t_0}{t}^{2} \right)\xi_0r{ \kappa}^{2} \right) r \right] \sqrt {r-r_0}\sqrt {rt_0+tr_0}+t_0 \left( t \left(  \left( \frac{16}3t{\kappa}^{2}\xi_0+\frac{4}3{\kappa}^{2}\xi_0t_0+t{t_0}^{2} \right) {r}^ {2}+\frac{8}3{t_0}^{3}{\kappa}^{2}\xi_0 \right) {r_0}^{2}\right.\right.\right.\nonumber\\
&\left.\left.\left. +2  \left[  \left( t{t_0}^{2}+\frac{2}3{\kappa}^{2}\xi_0t_0 +\frac{16}3t{\kappa}^{2}\xi_0 \right) {r}^{2}-\frac{10}3\xi_0{t_0}^{2}{\kappa}^{2} \left( t+\frac{t_0}5 \right)  rt_0 r_0+{r}^{2}{t_0}^{2} \left\}  \left( {t_0}^{2}+\frac{16}3{ \kappa}^{2}\xi_0 \right) {r}^{2}-\frac{8}3{t_0}^{2}{\kappa}^{2}\xi_0 \right\}  \right]  \left( r-r_0 \right)  \right){e}^{2 {\frac {t}{t_0}}}\right.\right.\nonumber\\
&\left.\left.  +6{r}^{2}{r_0}^{2} \left[ \frac{16}3 \left\{  {\frac {r_0}{32}} \left( 3 \left(2r^2+16\kappa^2\xi_0{}^2 \right) r_0-3{r}^{3}-40r{\kappa}^{2} \xi_0 \right) t{t_0}^{7/2}+ \left[  \left\{  \left( {\frac {3} {32}}{r}^{2}+\frac{7}4{\kappa}^{2}\xi_0 \right) r_0-\frac{3}16r \left( 8{\kappa}^{2}\xi_0+{r}^{2} \right)  \right\} {t_0}^{ 9/2}\right.\right.\right.\right.\right.\nonumber\\
&\left.\left.\left.\left.\left. + \left\{ 2 \left( r-\frac{r_0}8 \right) r_0t{t_0} ^{3/2}+ \left( r-\frac{r_0}4 \right) r{t_0}^{5/2}+{r_0}^{2 }\sqrt {t_0}{t}^{2} \right\} \xi_0r{\kappa}^{2} \right] r \right\} \sqrt {r-r_0}\sqrt {rt_0+tr_0}+t_0 \left\{ \left( \frac{4}3{\kappa}^{ 2}\xi_0t_0-\frac{16}3t{\kappa}^{2}\xi_0+t{t_0}^{2} \right) {r}^{2}\right.\right.\right.\right.\nonumber\\
&\left.\left.\left.\left.-8{t_0}^{3 }{\kappa}^{2}\xi_0 \right\} {r_0}^{2}+2r \left\{  \left( t{t_0}^{2}+\frac{2}3{\kappa}^{2}\xi_0t_0-\frac{16}3t{\kappa}^{2}\xi_0 \right) {r}^{2}+2{t_0}^{2}{\kappa}^{2}\xi_0 \left( t-3t_0 \right)  \right\} t_0r_0+{r}^{2} \left\{  \left( {t_0}^{2}-\frac{16}3{\kappa}^{2}\xi_0 \right) {r} ^{2}+8{t_0}^{2}{\kappa}^{2}\xi_0 \right\} {t_0}^{2} \right]\right.\right.\nonumber\\
&\left.\left.  \left( r-r_0 \right)  \right\} {e}^{4{\frac {t}{t_0}}}+2 \left( {r}^{6}{e}^{6{\frac {t}{t_0}}}+{r_0}^{6} \right)  \left\{  \left[ \frac{3}2 \left( \frac{r_0}3r_0-r \right) r_0t{t_0}^{7/2}-{t_0}^{9/2} \left( r-\frac{r_0}2 \right) r \right] \sqrt {r-r_0}\sqrt {rt_0+tr_0}+t_0^{3} \left( r-r_0 \right)  \left( rt_0+tr_0 \right) ^{2} \right\}  \right]\nonumber\\
&\left[ {t_0}^{3}{\kappa}^{2}{r}^{2} \left( rt_0+tr_0 \right) ^{2} \left( r-r_0 \right)\left( {r}^{2}{e}^{2{\frac {t}{t_0}}}+{r_0}^{2} \right) ^{3}\right]\,.
\end{align}

\end{document}